\documentclass[a4paper,11pt,bibtotoc,liststotoc]{article}

\usepackage{amssymb}
\usepackage{graphicx}
\usepackage{hyperref}
\usepackage[cal=esstix]{mathalfa}
\usepackage{amsmath}
\usepackage[english]{babel}
\usepackage{amsfonts}
\usepackage{rotating}
\usepackage{pdflscape}
\usepackage[official]{eurosym}
\usepackage{longtable}
\usepackage[left=2.5cm,right=2.5cm,top=2cm,bottom=2cm]{geometry}
\usepackage{xspace}
\usepackage{array}
\usepackage{latexsym}
\usepackage{bm}
\usepackage{bbm}
\usepackage{xcolor}
\usepackage{xcolor,colortbl}
\usepackage{color,soul}
\usepackage{algpseudocode}
\usepackage{algorithm}

\usepackage[flushleft]{threeparttable}
\usepackage{rotating}
\usepackage{tabularx}
\usepackage[section]{placeins}

\usepackage[toc,page]{appendix}

\usepackage{adjustbox}

\usepackage{xspace}
\usepackage{authblk}

\usepackage{booktabs}
\definecolor{darkgreen}{rgb}{0, 0.5, 0}

\usepackage[round,longnamesfirst]{natbib}
\usepackage{subfiles}
\usepackage{subcaption}

\newcommand{\indep}{\rotatebox[origin=c]{90}{$\models$}}

\begin{document}

\title{\Large{\textbf{Identifying Causal Effects of Discrete, Ordered and Continuous Treatments using Multiple Instrumental Variables}}}

\date{\today}
\author{Nadja van 't Hoff\footnote{Email address: navh@sam.sdu.dk.} \\ 
University of Southern Denmark}

\maketitle

\vspace{-0.25cm}
\begin{center}
    Click \href{https://nadjavanthoff.com/files/JMP_Nadja_vantHoff.pdf}{\textcolor{blue}{here}} for the most recent version.
\end{center}

\vspace{0.25cm}
\begin{abstract} \noindent
Inferring causal relationships from observational data is often challenging due to endogeneity. This paper provides new identification results for causal effects of discrete, ordered and continuous treatments using multiple binary instruments. 
The key contribution is the identification of a new causal parameter that has a straightforward interpretation with a positive weighting scheme and is applicable in many settings due to a mild monotonicity assumption. 
This paper further leverages recent advances in causal machine learning for both estimation and the detection of local violations of the underlying monotonicity assumption.
The methodology is applied to estimate the returns to education and assess the impact of having an additional child on female labor market outcomes.
 \bigskip 
 
\end{abstract}

\vfill

{\small \bigskip {\footnotesize \noindent \textit{This project was supported by generous funding from the Independent Research Fund Denmark (90380031B).
I am grateful to Giovanni Mellace for his guidance and to my fellow PhD students for their feedback.
I extend my thanks to Michael Lechner, Toru Kitagawa, Volha Lazuka, Guido Imbens, Bo Honor\'e, Martin Huber, Phillip Heiler, Leonard Goff, and Kevin Huynh for their insightful comments. 
I also appreciate the feedback from the participants at ``Machine Learning in Program Evaluation, High-Dimensionality, and Visualization Techniques'' 2023, IAAE 2023, the seminar at Lund University in October 2023, DGPE Workshop 2023, (EC)\^{}2 2023, EWMES 2023, RES Conference 2024, the seminar at UvA in March 2024, Aarhus Workshop in Econometrics III, COMPIE 2024, the ``Groningen Workshop on Causal Inference and Machine Learning'' 2024, and the seminar at the University of St. Gallen in October 2024.
}
\thispagestyle{empty}\pagebreak }}

{\small \renewcommand{\thefootnote}{\arabic{footnote}} %
\setcounter{footnote}{0} \pagebreak \setcounter{footnote}{0} \pagebreak %
\setcounter{page}{1} }

\newpage

\section{Introduction}

Identifying causal relationships is a central goal in economic research, but inferring causality from observational data is often challenging, particularly due to endogeneity arising from selection into treatment. Instrumental variable methods are widely used to address this issue.

Much of the existing literature focuses on identifying causal effects of binary treatments with a single instrument. Yet many real-world applications involve discrete or continuous treatments and multiple instruments can be available. For example, rather than estimating the effect of education using binary indicators like college completion, one might have data on a discrete, ordered treatment such as years of schooling. Moreover, multiple instruments might be available, such as quarter of birth \citep{angrist1995two}, distance to school, or local labor market conditions \citep{carneiro2011estimating}. When treatment effects are heterogeneous, combining multiple instruments is advantageous. Each instrument generates distinct complier populations, and combining them expands the overall complier population considered, potentially bringing the local effect closer to the average treatment effect (ATE).

Two-stage least squares (TSLS) has become the standard estimation approach, but it faces two key limitations when applied to cases involving treatments with variable intensity and multiple instruments. First, the TSLS estimand is complex: \citet{angrist1995two} show that it represents a weighted average of average causal responses (ACRs), where each ACR reflects the effect of a one-level treatment change for specific subpopulations. In essence, the TSLS estimand consists of weighted averages of weighted averages, complicating interpretation of the effect estimates. Second, TSLS relies on a restrictive monotonicity assumption, already in the case of binary treatments. \citet{mogstad2021causal} show that while relaxing this assumption preserves the TSLS interpretation as a weighted average, it introduces the possibility of negative weights, further complicating the effect interpretation. This issue parallels concerns raised with two-way fixed effects estimators \citep{borusyak2018revisiting,de2020two,goodman2021difference}, where more complex settings with relaxed assumptions result in estimands that are harder to interpret.

The main contribution of this paper is the identification of a novel causal parameter for discrete, ordered and continuous treatments with multiple instruments. This parameter features an intuitive, positive weighting scheme and is derived under a mild monotonicity assumption. Specifically, this paper introduces the \textit{combined compliers average causal response} (CC-ACR), identified under the limited monotonicity (LiM) assumption. The CC-ACR parameter is easy to interpret, with weights that reflect the proportion of combined compliers, a sizable complier group within the population. It addresses the question, ``What is the average causal effect of a one-level increase in treatment for combined compliers?'', making it easier to interpret than the TSLS estimand, while imposing a less restrictive form of monotonicity. 

When treatment effects are heterogeneous, the monotonicity assumption is critical for ensuring a causally interpretable effect. In general, the monotonicity assumption restricts the direction in which the potential treatment status changes for given changes in instrument values, effectively ruling out certain response types and imposing restrictions on choice mechanisms. 
The CC-ACR is identified under the LiM assumption, which was originally introduced for binary treatments by \citet{lewbel2022limited}. It imposes fewer restrictions on choice behavior than the Imbens and Angrist monotonicity (IAM) assumption \citep{angrist1995two} or the partial monotonicity (PM) assumption introduced by \citet{mogstad2021causal}. The latter was introduced in the context of binary treatments to relax the IAM assumption.

This study extends the concepts of PM and LiM to the framework of discrete, ordered or continuous treatments. Interestingly, in this context, PM shares similar limitations as IAM, which were highlighted by \citet{mogstad2021causal} in the binary context. I show that LiM, the monotonicity assumption underlying my main identification result, not only avoids these limitations but also reduces the multiple instrument problem to a single instrument framework. In addition, it allows for the grouping of complier types, simplifying interpretation.

An additional contribution of this study is a general TSLS identification result that flexibly incorporates the various monotonicity assumptions, illustrating how each assumption influences the TSLS estimand. The challenge with TSLS lies in navigating a landscape of suboptimal options. While IAM guarantees positive weights, it is restrictive with respect to choice behavior. In contrast, PM offers greater flexibility by relaxing these restrictions to some extent, but at the risk of introducing negative weights.
In both cases, the TSLS estimator converges to a weighted average of weighted averages of ACRs, and this weighting scheme complicates the interpretation of TSLS estimates.

Another key contribution of this paper is a stochastic dominance test for detecting violations of the LiM assumption. LiM implies positive weights for the CC-ACR parameter, which is equivalent to the condition that the cumulative distribution functions (CDFs) of the treatment, conditional on specific instrument values, do not intersect. This necessary (but not sufficient) condition can be tested empirically. Since violations of LiM in specific subgroups may average out in the full sample, a global test might fail to detect them. Building on \citet{farbmacher2022instrument}, I show how causal forests can be used to detect such local violations by checking the sign of conditional average treatment effects within regression tree leaves, where a positive sign indicates a violation. An advantage of this method is that it enables data-driven subgroup formation in the covariate space.

To demonstrate the proposed methods, I revisit two seminal applications. The first examines the returns to education as studied by \citet{card1995geographic}. Rather than focusing on the binary variable of college attendance, I focus on education measured as a discrete variable, defined by the number of years of schooling. To address potential unobserved confounding factors, such as ability, which may influence both education and wages, I follow \citet{card1995geographic} and use the presence of 2-year and 4-year colleges as instruments. The combined compliers in this context are individuals whose education level increases due to living near a 2-year or 4-year college or both. The CC-ACR parameter then captures the average causal effect of an additional year of schooling for these individuals, providing a clear interpretation that cannot be obtained through TSLS. Moreover, LiM, the monotonicity assumption required for identifying the CC-ACR, is generally more plausible than the monotonicity assumptions required for reasonable TSLS estimates. LiM accommodates individuals who prefer 2-year colleges as well as those who prefer 4-year colleges, while other assumptions do not.
From a policy standpoint, understanding the impact of an additional year of schooling, rather than merely focusing on college completion, offers valuable insights for various analyses. This broader perspective can inform a range of policy decisions, including whether to extend or shorten the duration of college attendance.

Following \citet{card1995geographic}, I assume the instruments are exogenous, conditional on covariates such as individual and regional characteristics. I establish an identification result for the CC-ACR under conditional instrument validity, using similar arguments as \citet{frolich2007nonparametric}. A minor contribution of this paper is the extension of recent advances in causal machine learning, specifically double/debiased machine learning (DML) \citep{chernozhukov2018double}, from the binary treatment and single instrument setting to one with a discrete, ordered or continuous treatment and multiple binary instruments. This estimation approach has the advantage of accommodating a larger number of covariates and their interactions compared to other nonparametric estimators, which is crucial to maintain the interpretation of the CC-ACR. However, estimation is conducted on a subset of the sample, which is the cost of obtaining both a more interpretable effect estimate and a more credible monotonicity assumption.

\citet{card1995geographic} primarily focuses on the 4-year college instrument, given the weakness of the 2-year college instrument. However, an advantage of the proposed methodology is that information from the weaker instrument can still be included, as weak instruments do not necessarily pose a problem when paired with a strong instrument. The results indicate that individuals attending 2-year colleges may especially benefit from additional schooling.

The second application explores the impact of an additional child on female labor market outcomes, measured as annual labor income, weekly hours worked, and weeks worked per year, following the study by \citet{angrist1996children}. Next to the classical same-sex instrument, I use a twinning instrument that accounts for twins at any birth. In this application, LiM is more plausible than other forms of monotonicity, as it accommodates individuals who prefer same-sex as well as those who prefer mixed-sex siblings. A key advantage of the CC-ACR is that it provides insights into the average effect of having an additional child, rather than focusing solely on the effect of a third child. I find relatively small effects of an additional child on female labor market outcomes, likely due to the underlying complier population. Women with more children may have different labor preferences, which may explain the modest impact.

The remainder of this paper is organized as follows: Section \ref{sec:literature} reviews the relevant literature. Section \ref{sec:framework_and_identification} outlines the framework, assumptions, and the CC-ACR and TSLS identification results. Section \ref{sec:testing_lim} details the procedure for detecting violations of the LiM assumption, while Section \ref{sec:estimation} provides guidelines for estimation. Section \ref{sec:results} presents the empirical findings from the two seminal studies \citep{card1995geographic,angrist1996children}. Finally, Section \ref{sec:discussion} offers a discussion and suggests avenues for future research. Additional results, including simulation studies for the proposed LiM test, are provided in the appendix. The code used in this study will be made publicly available on GitHub.

\bigskip

\section{Literature review}
\label{sec:literature}

This paper contributes to the instrumental variables literature in two key areas. First, it enhances our understanding of which causal parameters can be identified using instrumental variables. The foundation of the local average treatment effect (LATE) framework was established by \citet{IA1994} and \citet{angrist1996identification}. For treatments with variable intensity, \citet{angrist1995two} show that TSLS combines the instrument-specific weighted averages into a new weighted average. However, most literature focuses on the case of a binary treatment.
In the setting with a binary treatment and multiple instruments, \citet{mogstad2021causal} provide an identification result for TSLS under their Partial Monotonicity (PM) assumption, and \citet{lewbel2022limited} show that the LATE for the combined compliers is identified under their Limited Monotonicity (LiM) assumption. \citet{goff2020vector} introduces Vector Monotonicity (VM), a special form of PM, which assumes that treatment uptake is monotonic with respect to each individual instrument, rather than requiring a uniform direction of response across all instruments. He further characterizes the class of causal parameters that are point-identified under this monotonicity assumption and provides a practical two-step estimator. \citet{frolich2007nonparametric} extends the LATE framework to include covariates nonparametrically. My findings complement those of \citet{frolich2007nonparametric}, who separately considers a discrete, ordered treatment with a single instrument, or a binary treatment with multiple instruments, while my results consider the setting with discrete, ordered treatments and multiple instruments.
My paper exhibits some connection to the work of \citet{lee2018identifying}, who study discrete treatments and the point-identification of marginal treatment effects (MTE) in a framework that requires continuous instruments. Equally within the MTE framework, \citet{heckman2006understanding} consider an ordered choice model, identifying a parameter for the difference in potential outcomes between two subsequent treatment levels. Unlike the approach in the present paper, their approach requires an instrument for all incremental changes in the treatment level. \citet{bhuller20222sls} extend \citet{frandsen2023judging}'s results for a binary treatment to a setting with multivalued treatments. They also require an instrument for every treatment level and assume no cross-effects, which is a rather restrictive assumption. Moreover, their result separately compares causal effects on specific treatment margins and does not offer an interpretation as an average effect for a one-level increase. For readers interested in a deeper exploration of instrumental variables methods that account for unobserved heterogeneity in treatment effects, I recommend the comprehensive review by \citet{mogstad2024instrumental}.

Second, this paper contributes to the literature on specification tests of instrument validity. Research in this area has mainly been limited to joint tests on the exclusion restriction and monotonicity for a binary treatment. The first testable implications based on the exclusion and monotonicity assumptions can be traced back to \citet{balke1997bounds}, \citet{angrist1995two}, and \citet{heckman2006understanding}. \citet{angrist1995two} show that, in the case of treatments with variable intensity and a single instrument, testable implications of IAM can be established. The testable implications in the present study consider the setting with multiple instruments and the LiM assumption.
There is a large body of literature that derives results related to testable implications for joint tests in the case of a binary treatment \citep{kitagawa2021identification,balke1997bounds,kitagawa2015test,mourifie2017testing,mellace2015testing,frandsen2023judging,carr2021testing}. \citet{farbmacher2022instrument} build on this literature, but employ causal forests to detect local violations of the joint assumptions, subgrouping the covariates in a data-driven way. The present paper complements this paper, as it provides a test using causal forests for LiM when the treatment is discrete and ordered. While the aforementioned literature focuses on a binary treatment, there has been some recent progress in extending the testable implications to non-binary treatment settings. For instance, \citet{sun2020instrument} establishes testable implications for the exclusion and IAM assumptions for ordered and unordered nonbinary treatments.

\bigskip

\section{Identification results}

\label{sec:framework_and_identification}

\subsection{Framework and assumptions}
\label{sec:framework_and_assumptions}

Consider the \citet{angrist1995two} setup with an outcome $Y$, a treatment $D$ that is discrete with bounded support, $D \in \{ 0,1,...,J \}$, such that there are $J+1$ possible treatment levels, and $K$ binary instruments, $Z_1$, $Z_2$, ..., and $Z_K$. Adhering to the Rubin causal model (as detailed in \citeauthor{rubin1974estimating}, \citeyear{rubin1974estimating}, and \citeauthor{robins1986a}, \citeyear{robins1986a}), the potential treatment states for some unit $i$ are denoted as $D_{i}^{z_{1}z_{2}...z_{K}}$, while potential outcomes are represented by $Y_{i}^{j,z_{1}z_{2}...z_{K}}$.

\bigskip
\noindent\textbf{Assumption 1: Random assignment and exclusion}\\
\indent $Z_{k} \indep (D^{z_{1}z_{2}...z_{K}},Y^{j}) \indent \forall z_{1}z_{2}...z_{K}\in\{0,1\}^K,k\in \{1,2,...,K\},j\in\{0,1,...,J\}$.

\bigskip
\noindent \textbf{Assumption 2: Stable unit treatment value assumption (SUTVA)} \\
\indent $Y_{i}^{j,z_{1}z_{2}...z_{K}}=Y^j$ and $Y=Y^j$ if $D=j$, and \\
\indent $D=D^{z_{1}z_{2}...z_{k}}$ if $Z_1=z_1$, $Z_2=z_2$, ..., and $Z_K=z_K$.

\bigskip
\noindent \textbf{Assumption 3: Instrument relevance} \\
\indent $0<P(Z_1 \cdot Z_2 \cdot ... \cdot Z_K = 1)<1$, and $0<P((1-Z_{1})\cdot(1-Z_{2})\cdot...\cdot(1-Z_{K}) = 1) <1 $, and \\ 
\indent $P(D^{1...1...1} \geq j > D^{0...0...0}) > 0$ for some $j\in \{0,1,...,J\}$.

\bigskip
\noindent \textbf{Assumption 4: Limited monotonicity (LiM)} \\
\indent $P(D^{1...1...1}\geq D^{0...0...0}) = 1$ \text{ or } $P(D^{1...1...1}\leq D^{0...0...0}) = 1$.
\bigskip

The validity of the instruments relies on the independence assumption and exclusion restriction, both outlined in Assumption 1. In \citeauthor{card1995geographic}'s (\citeyear{card1995geographic}) application, the independence assumption posits that the presence of a college does not influence an individual's wage other than through the change in years of schooling attained. 
SUTVA (Assumption 2) ensures that the treatment level of one unit remains unaffected by the treatment level of any other unit, and that instruments assigned to a specific unit solely impact the treatment level for that particular unit. SUTVA guarantees the existence of a singular potential outcome for each treatment value.
Assumption 3 requires the instruments to be relevant, which is important for estimation and for the existence of a complier population. This means that at least one instrument affects some level of the treatment to ensure the existence of compliers. For instance, this implies that the proximity to a college influences educational attainment for some individuals.

The limited monotonicity (LiM) assumption was initially introduced by \citet{lewbel2022limited} for the setting with a binary treatment. Assumption 4 extends LiM to settings where treatment intensity varies. It states that when exposed to all (none) of the instruments, units are at least as likely to take up treatment as when exposed to none (all) of the instruments simultaneously. This introduces restrictions on choice behavior at the outer support of the instrument values. Without loss of generality, positive LiM ($P(D^{1...1...1}\geq D^{0...0...0}) = 1$) is assumed throughout the rest of the paper. In \citeauthor{card1995geographic}'s (\citeyear{card1995geographic}) study, this implies that an individual's educational attainment while residing close to both a 2-year college and a 4-year college is at least as large as the number of months when residing far from both a 2-year college and a 4-year college.

LiM is generally weaker than other monotonicity assumptions introduced in the literature, such as the Imbens and Angrist monotonicity (IAM) assumption \citep{IA1994} and the partial monotonicity (PM) assumption \citep{mogstad2021causal}. Although Vector Monotonicity (VM) as introduced by \citet{goff2020vector} is not further discussed here, as it is a special case of PM, it is important to note that it is the most empirically relevant case of PM.

IAM evaluates potential treatment states for all instrument values, basically requiring individuals to prefer one instrument over another. On the other hand, PM restricts the direction of the potential treatment status for a change in one of the instruments while keeping all other instrument values fixed. While PM has been primarily been introduced for the setting with a binary treatment, this paper extends it seamlessly to the nonbinary treatment scenario.
\bigskip

\noindent\textbf{Imbens and Angrist monotonocity (IAM)} \\
\indent $P(D^{i...j...k} \geq D^{p...q...r}) = 1$ or $P(D^{i...j...k} \leq D^{p...q...r}) = 1 $ \\
\indent $\forall \ i \in \{0,1\}, ..., j \in \{0,1\}, ..., k \in \{0,1\}$ and $\forall \ p \in \{0,1\}, ..., q \in \{0,1\}, ..., r \in \{0,1\}$ \\
\indent such that $P(D^{i...j...k}) \neq P(D^{p...q...r})$.
\bigskip

\noindent\textbf{Partial monotonicity (PM)} \\
\indent $P(D^{1...j...k} \geq D^{0...j...k}) = 1$ or $P(D^{1...j...k} \leq D^{0...j...k}) = 1$, \\
\indent $P(D^{i...1...k} \geq D^{i...0...k}) = 1$ or $P(D^{i...1...k} \leq D^{i...0...k}) = 1$, and \\
\indent $P(D^{i...j...1} \geq D^{i...j...0}) = 1$ or $P(D^{i...j...1} \leq D^{i...j...0}) = 1$ \\
\indent $\forall \ i \in \{0,1\}, ..., j \in \{0,1\}, ..., k \in \{0,1\}$.
\bigskip

The restrictions on the choice mechanisms imposed by LiM are reflected in the response types of the population. In the case of a binary instrument and binary treatment, \citet{IA1994} introduce the notions of always-takers, compliers, defiers, and never-takers. Here, LiM rules out the defiers. 
Now consider the scenario with a three-valued treatment, $D \in \{ 0,1,2\}$, and one binary instrument, $Z \in \{ 0,1 \}$. There are $(J+1)^{2^K}=3^{2^1}=9$ initial response types. Adapting \citet{frolich2007nonparametric}'s notation, the non-responders, whose treatment level does not change in response to a change in the instrument ($D^1=D^0$), are denoted by $n_{{D^1},{D^0}}$. The compliers are the types denoted by $c_{{D^1},{D^0}}$ for whom $D^1>D^0$, while for defiers, $d_{{D^1},{D^0}}$, it holds that $D^1<D^0$. 
Compliers are individuals for which $P(D^1 \geq D^0) = 1$, while defiers have $P(D^0 \geq D^1) = 1$. Monotonicity rules out three defier types (see Table \ref{tab:types_one_inst}).

The number of initial response types increases rapidly with the number of treatment levels. For discrete, ordered and continuous treatments, compliance intensity can vary. This means that, for a certain change in the instrument values, some response types might shift their treatment status by one level, while others might shift their treatment status by two levels. In addition, these types can have distinct baseline treatment levels, $Y_i^0$, adding to the complexity of types. 

\begin{table}[!tbp]
    \centering
    \caption{Initial response types with one binary instrument, $Z \in \{ 0,1 \}$, and a three-valued treatment, $D \in \{ 0,1,2\}$. \checkmark indicates the response types allowed for under the different forms of the monotonicity assumption.}
    \label{tab:types_one_inst}
    \begin{tabular}{c|cc|ccc}
        \hline 
        Type & $D^0$ & $D^1$ & LiM & PM & IAM \\ \hline
        $c_{0,1}$ & 0 & 1 & \checkmark & \checkmark & \checkmark \\
        $c_{1,2}$ & 1 & 2 & \checkmark & \checkmark & \checkmark \\
        $c_{0,2}$ & 0 & 2 & \checkmark & \checkmark & \checkmark \\
        $n_{2,2}$ &2 & 2 & \checkmark & \checkmark & \checkmark \\
        $n_{1,1}$ & 1 & 1 & \checkmark & \checkmark & \checkmark \\
        $n_{0,0}$ & 0 & 0 & \checkmark & \checkmark & \checkmark \\
        $d_{1,0}$ & 1 & 0 & & & \\
        $d_{2,1}$ & 2 & 1 & & & \\
        $d_{2,0}$ & 2 & 0 & & & \\
         \hline
    \end{tabular}
\end{table}

Next consider the scenario involving a three-valued treatment, $D \in \{ 0,1,2\}$, while introducing two binary instruments, $Z_1  \in \{ 0,1 \}$ and $Z_2 \in \{ 0,1 \}$. This amplifies the number of potential response types to $(J+1)^{2^K}=3^{2^2}=81$. Refer to Appendix \ref{app:full_table_three_valued}, Table \ref{tab:all_types_3_values}, for a comprehensive listing of all initial response types. Under LiM, 54 response types remain, as types that defy with respect to the outer instrument support are eliminated, specifically types $d_{D^{00},...,D^{11}}$ where $D^{00} > D^{11}$.
Under PM with ordering $P(D^{10}\geq D^{00})=1$, $P(D^{01}\geq D^{00})=1$, $P(D^{01}\geq D^{11})=0$, $P(D^{10}\geq D^{11})=0$, a total of 20 response types remain. Under IAM, there are only 14 response types that remain.\footnote{For a detailed comparison of the three monotonicity assumptions when the treatment is binary see \citet{lewbel2022limited}.} IAM only allows for pure compliers in the sense that there are no two-way flows for any shift in the instrument values.
Altogether, LiM allows for more response types and hence for rich choice heterogeneity.

\begin{table}[!p]
    \centering
    \caption{This table presents the response types that are contained in the combined complier type $cc_{0,1}$. \checkmark indicates the response types allowed for under the different forms of the monotonicity assumption.}
    \label{tab:ex_c10}
    \begin{tabular}{ll|cccc|ccc}
    \hline
    Combined type & Type & $D^{00}$ & $D^{01}$ & $D^{10}$ & $D^{11}$ & LiM & PM & IAM \\
    \hline
    $cc_{0,1}$ & $c_{0,2,2,1}$ & 0 & 2 & 2 & 1 & \checkmark &  &  \\
     & $c_{0,1,2,1}$ & 0 & 1 & 2 & 1 & \checkmark &  &  \\
     & $c_{0,0,2,1}$ & 0 & 0 & 2 & 1 & \checkmark &  &  \\
     & $c_{0,2,1,1}$ & 0 & 2 & 1 & 1 & \checkmark &  &  \\
     & $c_{0,1,1,1}$ & 0 & 1 & 1 & 1 & \checkmark & \checkmark & \checkmark \\
     & $c_{0,0,1,1}$ & 0 & 0 & 1 & 1 & \checkmark & \checkmark & \checkmark \\
     & $c_{0,2,0,1}$ & 0 & 2 & 0 & 1 & \checkmark &  &  \\
     & $c_{0,1,0,1}$ & 0 & 1 & 0 & 1 & \checkmark & \checkmark &  \\
     & $c_{0,0,0,1}$ & 0 & 0 & 0 & 1 & \checkmark & \checkmark &  \\ \hline
    \end{tabular}
\end{table}

\begin{table}[!p]
    \centering
    \caption{All possible initial combined response types with two instruments, $Z_1 \in \{ 0,1 \}$ and $Z_2 \in \{ 0,1 \}$, and a three-valued treatment, $D \in \{ 0,1,2\}$. \checkmark indicates the response types allowed for under LiM. Combined compliers are denoted by $cc_{D^{00},D^{11}}$, combined non-responders by $cn_{D^{00},D^{11}}$, and combined defiers by $cd_{D^{00},D^{11}}$.}
    \label{tab:aggregated_types_three_valued}
    \begin{tabular}{l|cc|c} \hline
        Combined type & $D^{00}$ & $D^{11}$ & LiM \\ \hline
         $cc_{0,1}$ & 0 & 1 & \checkmark \tabularnewline
         $cc_{1,2}$ & 1 & 2 & \checkmark \tabularnewline
         $cc_{0,2}$ & 0 & 2 & \checkmark \tabularnewline
        $cn_{2,2}$ & 2 & 2 & \checkmark \\
        $cn_{1,1}$ & 1 & 1 & \checkmark \\
        $cn_{0,0}$ & 0 & 0 & \checkmark \\ 
        $cd_{2,0}$ & 2 & 0 & \\
        $cd_{2,1}$ & 2 & 1 &  \\ 
        $cd_{1,0}$ & 1 & 0 & \\ \hline
    \end{tabular}
\end{table}
 
In addition to allowing for rich choice heterogeneity, LiM allows us to aggregate the response types into groups, reducing the complex problem with many response types to a simpler one.
Since LiM only imposes a restriction on the outer support ($Z_1=Z_2=1$ and $Z_1=Z_2=0$) of $\mathcal{Z}=\{ (1,1), (1,0), (0,1), (0,0)\}$, the two intermediate treatment states, $D^{10}$ and $D^{01}$, are not restricted. Therefore, aggregating the initial response types into response type groups is straightforward. 
With two instruments, I define as combined compliers, denoted as $cc_{D^{00},D^{11}}$, those types who increase the treatment level in response to changing both instrument values from zero to one ($D^{11}>D^{00}$), combined defiers, denoted as $cd_{D^{00},D^{11}}$, those types who have $D^{11}<D^{00}$, and as combined non-responders, denoted as $cn_{D^{00},D^{11}}$, those types who have $D^{11}=D^{00}$.  
It is important to highlight that aggregating the groups in this way is not possible under PM or IAM.

To illustrate this, consider the nine initial types in Table \ref{tab:ex_c10}, extracted from Table \ref{tab:all_types_3_values} in Appendix \ref{app:full_table_three_valued}. These nine types can be aggregated into a single combined complier type, recognizing that shifting the instrument values from $(0,0)$ to $(1,1)$ increases the potential treatment status from zero to one across all nine types. This combined complier type can be denoted as $cc_{0,1}$. At the intermediate instrument values, namely $(1,0)$ and $(0,1)$, this aggregated type can respond as complier or defier with respect to either instrument. In a similar fashion, the remaining response types in Table \ref{tab:all_types_3_values} can be aggregated into groups, effectively reducing the initial 81 types to the nine aggregated types showcased in Table \ref{tab:aggregated_types_three_valued}. Similar results apply to settings where the treatment attains more than three levels or where more than two binary instruments are available. LiM naturally reduces a complex setting to a simple comparison between two different potential treatment states, $D^{1...1...1}$ and $D^{0...0...0}$, independently of the number of instruments. Notably, combined defier types with $cd_{a,b}$ where $b<a$ are ruled out by the LiM assumption, but all other defier types are not.

\bigskip

\subsection{The combined compliers ACR}
\label{sec:main_results}

Theorem 1 provides the main result, namely the CC-ACR, a novel causal parameter which has a straightforward interpretation and is derived under the LiM assumption.

\bigskip
\noindent \textbf{Theorem 1: The combined compliers average causal response (CC-ACR) \\} 
\textit{Let Assumptions 1\footnote{Assumption 1 can be relaxed to hold only for the instrument $\widetilde{Z}$, where $\widetilde{Z}=1$ if $Z_1=Z_2=\dots=Z_K=1$ and $\widetilde{Z}=0$ if $Z_1=Z_2=\dots=Z_K=0$, but the original form of Assumption 1 is equally plausible in most cases.}, 2, 3, and 4 hold. Then a weighted average of average causal responses for the combined complier subpopulations is identified:}

\begin{equation}\label{eq:thm_1}
    \begin{aligned}
        \beta_{\text{CC-ACR}} &\equiv \frac{E(Y|Z_1=Z_2=...=Z_K=1) - E(Y|Z_1=Z_2=...=Z_K=0)}{E(D|Z_1=Z_2=...=Z_K=1) - E(D|Z_1=Z_2=...=Z_K=0)} \\
        &= \sum_{k < l} \frac{(l-k) \cdot P(T= cc_{k,l})}{\sum_{m < h} (h-m) \cdot P(T = cc_{m,h})} \cdot E\left(\frac{Y^l - Y^k}{l-k}|T=cc_{k,l}\right).
    \end{aligned}
\end{equation}
\bigskip

\noindent \textbf{Proof} in Appendix \ref{app:proof_theorem1}. \bigskip

$T$ denotes type and the set of response types, $cc$, consists of the combined complier types denoted $cc_{k,l}$ where $l>k$. These are the complier types that increase their treatment level in response to shifting all instruments from zero to one. 
Theorem 1 states that a weighted average of causal responses, $E(Y^l-Y^k)$, that are scaled by the change in treatment level, $(l-k)$, over these combined complier subpopulations is identified. It should be noted that this identification result is robust to the presence of non-responders.
It is further important to emphasize that the weights of the CC-ACR are always positive by construction and sum up to one.

Theorem 1 takes into account that the treatment responses vary in intensity. 
Within the context of the returns to education as considered by \citet{card1995geographic}, the CC-ACR provides a weighted average of causal responses for individuals who extend their education when all instrument values shift from zero to one. In this case, all instrument values equaling one indicates that individuals reside close to both a 2-year and a 4-year college.
To clarify Theorem 1, note that $E(Y^{14} - Y^{12} \mid cc_{12,14})$ measures the average effect on an individual's wage when obtaining 14 instead of 12 years of schooling, for those who adjust their education level accordingly in response to this change in instrument values. This average effect is weighted by the probability of belonging to this complier type, represented by $P(T = cc_{12,14})$, providing weights proportional to the response type group size.
Further, $E(Y^{14} - Y^{12} \mid cc_{12,14})$ represents the effect of 2 additional years of schooling, reflected in scaling the difference in outcomes, $Y^{14} - Y^{12}$, by the treatment level difference, $(l - k) = 2$ years.

Theorem 1 simplifies if treatment effects are linear, meaning the impact of increasing schooling by one year is equivalent, whether it is from 12 to 13 years or from 15 to 16 years.
In this case, the CC-ACR can be interpreted as the average effect of a one-level increase among the combined complier population: \(E(Y^j - Y^{j-1} \mid T \in cc)\), without considering the treatment margins involved. The following corollary demonstrates the interpretation of treatment effects within this linear framework.

\bigskip
\noindent \textbf{Corollary 1: Linearity of treatment effects}  \\
\textit{Let Assumptions 1, 2, 3, and 4 hold. Under linearity of the treatment effects, it holds for every treatment level, $j \in {1,...,J}$, that}
\begin{equation}\label{eq:thm_1_linear}
    \begin{aligned}
        \beta_{\text{CC-ACR}} &\equiv \frac{E(Y|Z_1=Z_2=...=Z_K=1) - E(Y|Z_1=Z_2=...=Z_K=0)}{E(D|Z_1=Z_2=...=Z_K=1) - E(D|Z_1=Z_2=...=Z_K=0)} \\
        &= \sum_{k < l} \frac{(l-k) \cdot P(T= cc_{k,l})}{\sum_{m < h} (h-m) \cdot P(T = cc_{m,h})} \cdot E\left(Y^{j} - Y^{j-1}|T=cc_{k,l}\right).
    \end{aligned}
\end{equation}
\bigskip

\noindent The following illustrative example clearly shows how Corollary 1 emerges from Theorem 1:
\begin{align*}
    & E\left(\frac{Y^2 - Y^0}{2-0}|T=cc_{0,2}\right) = E\left(\frac{Y^1 - Y^0}{2}|T=cc_{0,2}\right) + E\left(\frac{Y^2 - Y^1}{2}|T=cc_{0,2}\right) \\ &= 2 \cdot E\left(\frac{Y^1 - Y^0}{2}|T=cc_{0,2} \right)=E(Y^1 - Y^0|T=cc_{0,2})=E(Y^2 - Y^1|T=cc_{0,2}).
\end{align*}

\noindent This example illustrates that under linear treatment effects, the expected difference in the outcome when changing the treatment status from zero to one is equivalent to the expected difference when changing the treatment status from one to two for the combined compliers of type $cc_{0,2}$. These are the response types that would change their treatment status from zero to two when all instruments are changed from zero to one.


\bigskip

\subsection{Identification of the CC-ACR including covariates}
\label{sec:identification_covariates}

The result presented in the preceding section did not consider identification in the presence of relevant covariates. However, numerous real-world applications exist where the instruments are only valid after conditioning on covariates.
Taking the returns to education application as an example, one might worry about factors influencing an individual's surroundings and their wage. The presence of a college might correlate with various individual and county characteristics. Thus, Assumption 1 must be adjusted so that the instruments are approximately randomly assigned conditional on these characteristics.

\bigskip
\noindent \textbf{Assumption 1C: Unconfoundedness and exclusion} \\
\indent $Z_{k}\bot (D^{z_{1}z_{2}...z_{K}},Y^{j})|X \indent \forall z_{1},z_{2},...,z_{K},k\in \{1,2,...,K\},j\in\{0,1,...,J\}$.
\bigskip

In addition to Assumption 1C and Assumptions 2 to 4, common support is assumed to guarantee that there is overlap in the observed characteristics at the outer support of the instrument distribution.

\bigskip
\noindent\textbf{Assumption 5: Common support} \\
\indent $P(X=x|Z_1=z_1,Z_2=z_2,...,Z_K=z_k) > 0 \indent \forall x \in \mathcal{X}, \forall z_{1}z_{2}...z_{K}\in\{0,1\}^K$.
\bigskip

\noindent Theorem 1 can easily be extended to hold conditional on covariates:

\begin{equation}\label{eq:thm1_x}
        \begin{aligned}
            \beta_{\text{CC-ACR}}(X) &= \frac{E(Y|X,Z_1=Z_2=...=Z_K=1) - E(Y|X,Z_1=Z_2=...=Z_K=0)}{E(D|X,Z_1=Z_2=...=Z_K=1) - E(D|X,Z_1=Z_2=...=Z_K=0)} \\
            &= \sum_{k < l} \frac{(l-k) \cdot P(T= cc_{k,l}|X)}{\sum_{m < h} (h-m) \cdot P(T = cc_{m,h})} E\left(\frac{Y^l - Y^k}{l-k}|X,T=cc_{k,l}\right).
        \end{aligned}
\end{equation}
\bigskip

Then, to obtain $\beta_{\text{CC-ACR}}$, integrating over the distribution function $f_{x|\text{combined complier}}(x)$ is required. This function is unknown, but following \citet{frolich2007nonparametric} and using Bayes' theorem, it is straightforward to show that $f_{x|\text{combined complier}}(x)$ equals the estimable distribution function $f_x$, weighted with the corresponding increments of in the treatment level, $(l-k)$:

\begin{equation*}
    f_{x|\text{combined complier}}(x) = \frac{\sum_k^K \sum_{k<l}^K P(T=cc_{k,l}|X)\cdot(l-k)}{\sum_k^K \sum_{k<l}^K P(T=cc_{k,l})\cdot(l-k)} \cdot f_x(x).
\end{equation*} 
\bigskip

\noindent This allows for identification of the CC-ACR under Assumption 1C, as formalized in Corollary 2.

\noindent \textbf{Corollary 2: The CC-ACR under unconfoundedness} \\
\textit{Let Assumptions 1C and 2 to 5 hold. Then, the CC-ACR is given by}
\begin{equation*}
    \begin{aligned}
        & \beta_{\text{CC-ACR}} \\ &= \int \beta(x) \cdot f_{x|\text{combined complier}}(x)dx \\
        &= \frac{ \int (E(Y|X=x,Z_1=Z_2=...=Z_K=1)-E(Y|X=x,Z_1=Z_2=...=Z_K=0)) \cdot f_x(x) dx}{\int (E(D|X=x,Z_1=Z_2=...=Z_K=1)-E(D|X=x,Z_1=Z_2=...=Z_K=0)) \cdot f_x(x) dx}.
    \end{aligned}
\end{equation*}
\bigskip

For brevity, define 
\begin{align*}
    \widetilde{Z} =
    \begin{cases} 
        1 & \text{ if } Z_1=Z_2=...=Z_K=1 \\
        0 & \text{ if } Z_1=Z_2=...=Z_K=0
    \end{cases}.
\end{align*} 
Considering only the subsample at the outer support of the instrument distribution with $\widetilde{Z}$ as the sole instrument, this expression reduces to

\begin{equation}\label{eq:cc_acr_covariates}
    \beta_{\text{CC-ACR}} = \frac{ \int (E(Y|X=x,\widetilde{Z}=1)-E(Y|X=x,\widetilde{Z}=0)) \cdot f_x(x) dx}{\int (E(D|X=x,\widetilde{Z}=1)-E(D|X=x,\widetilde{Z}=0)) \cdot f_x(x) dx}.
\end{equation}
\bigskip

\subsection{The causal interpretation of two-stage least squares}

In this section, I extend prior research by \citet{mogstad2021causal}, which primarily delves into the causal interpretation of two-stage least squares (TSLS) with a focus on a binary treatment and multiple, mutually-exclusive instruments under the PM assumption. The goal is to generalize this result to the broader context of a discrete, ordered or continuous treatment, without imposing monotonicity at first. The probability limit of TSLS is given by Proposition 1. 


\bigskip
\noindent \textbf{Proposition 1: The causal interpretation of TSLS} \\
\textit{Let $M$ denote the number of elements in the rectangular instrument support $\mathcal{Z}=\{ z_0, ...,z_l, ..., z_m\}$, ordered such that $l<m$ implies $E(D|Z=l)<E(D|Z=m)$. Let $I(\cdot)$ denote the indicator function, which equals one if its argument is true and zero otherwise. Suppose that Assumptions 1, 2, and 3 are satisfied. Then,}
\begin{equation}\label{eq:beta_tsls}
    \begin{aligned}
        \beta_{\text{TSLS}} &= \sum_{t \in \mathcal{T_\text{M}}} P(T=t)  \sum_{m=1}^M \iota_{m,m-1} \cdot \omega_m \cdot E(Y^{D^{z_m}}-Y^{D^{z_{m-1}}}|T=t), 
    \end{aligned}
\end{equation}
\textit{where }
\begin{equation*}
    \begin{aligned}
        \omega_m &= \frac{(1-P(Z\geq z_m)) P(Z \geq z_m) \cdot \left\{ E(D|Z\geq z_m) - E(D|Z<z_m) \right\}}{\sum_{l=0}^M P(Z=z_l) E(D|Z=z_l)(E(D|Z=z_l)-E(D))},
    \end{aligned}
\end{equation*}
\textit{and}
\begin{equation*}
  \iota_{m,m-1} \equiv I(D^{z_m}\geq D^{z_{m-1}}) - I(D^{z_m}\leq D^{z_{m-1}}),      
\end{equation*}
\textit{where }$\mathcal{T_\text{M}}$ \textit{is the set of response types that are allowed for under the specified monotonicity assumption}. \bigskip

\noindent \textbf{Proof} in Appendix \ref{app:proof_tsls}.
\bigskip

Proposition 1 reveals that TSLS gives a weighted average of average causal responses (ACR), $E(Y^{D^{z_m}}-Y^{D^{z_{m-1}}})$, corresponding to the response types, $t$, present in the population. The weights determine the contribution of each local average causal response to the parameter $\beta_\text{TSLS}$.
Similar to the CC-ACR, the weights consist of $P(T=t)$, the probability of observing a certain response type, and are non-negative and sum to one.
However, the TSLS estimand contains additional, rather arbitrary weighting terms.
Consider, for instance, the weights $\omega_m$. 
These weights are proportional to $P(Z\geq z_m) (1-P(Z\geq z_m))$, effectively giving more weight to $E(Y^{D^{z_m}}-Y^{D^{z_{m-1}}}|T=t)$ when it lies in the center of the instrument distribution. 
It is hard to come up with an empirical setting where this is a desirable feature of the TSLS weights.
Ordering the values of the instrument support $\mathcal{Z}=\{ z_0, ...,z_l, ..., z_m\}$, such that $l<m$ implies $E(D|Z=l)<E(D|Z=m)$, results in $(E(D|Z \geq z_m ) - E(D|Z<z_m))$ being positive for all $z_m$. Note that this implies that the constructed instrument should be monotonic with the propensity score to ensure non-negative weights $\omega_m$.
This expression shows that more weight is given if comparatively more types respond to a change in the instrument values.  Next to the weight $\omega_m$, the TSLS estimand contains the term $\iota_{m,m-1}$, which can attain three values: $\iota_{m,m-1}$ equals 1 when $D^{z_m} > D^{z_{m-1}}$, it equals -1 when $D^{z_m} < D^{z_{m-1}}$, and 0 when $D^{z_m} = D^{z_{m-1}}$. In the latter case, it holds that $E(Y^{D^{z_m}}-Y^{D^{z_{m-1}}}|T=t)=0$. $\iota_{m,m-1}$ guarantees the interpretation of a weighted average of causal responses $Y^a-Y^b$ for which $a>b$. Simply put, it switches $E(Y^{D^{z_m}}-Y^{D^{z_{m-1}}})$ to $E(Y^{D^{z_m-1}}-Y^{D^{z_{m}}})$ whenever $D^{z_m-1}>D^{z_m}$.

Proposition 1 is derived without imposing any form of the monotonicity assumption. 
The advantage is that it enables researchers to reflect upon the response types existing in the population, drawing on prior knowledge or subject expertise.
Proposition 1, therefore, provides flexibility in formulating monotonicity.
Without imposing monotonicity, $\beta_\text{TSLS}$ contains negative weights because $\iota_{m,m-1}$ is always less than or equal to 1 for some response types. This results in a negative weight on the local ACRs for these types, reversing the sign of these specific ACRs in the weighted average of ACRs. Imposing certain forms of monotonicity can eliminate those types for which sign reversals in the average local effects occur.
For instance, IAM only allows for types which never have $\iota_{m,m-1}=-1$, thus guaranteeing positive weights. However, IAM does have the drawback of ruling out numerous response types, thereby heavily constraining choice heterogeneity. 
PM is often more reasonable to assume, given that it allows for a broader range of response types.
Under PM, response types are allowed to be present even when $\iota_{m,m-1}=-1$ for some $m$, provided that they have $\iota_{m',m'-1}=1$ for some other $m'$.
That is, PM ensures that the allowed response types also respond with an increase in the treatment level for some change in instrument values such that the local causal responses in the expression $\sum_{m=1}^M \iota_{m,m-1} \cdot \omega_m \cdot E(Y^{D^{z_m}}-Y^{D^{z_{m-1}}}|T=t)$ for this type $t$ are not all negatively weighted. 
It is alarming that, if the defier responses outweigh the complier responses,
the ACR for this type is negatively weighted. 

In short, researchers face a choice between imposing the more restrictive IAM assumption, which guarantees positive weights, or adopting PM with the risk of introducing sign-reversal issues through the weights $\iota_{m,m-1}$. In either scenario, the weights $\omega_m$ remain arbitrary and lack intuition. Moreover, PM might still be too restrictive in certain applications.
Notably, the CC-ACR parameter outlined in Theorem 1 bypasses these concerns: its weights are  positive by construction and it is identified under the generally weaker LiM assumption. Note that under LiM, the TSLS estimand will always include some negative weights.

\bigskip

\subsection{Alternative representation of the CC-ACR}

Theorem 1 above introduced the CC-ACR, expressed as a weighted average of causal responses for different response types. This representation is extremely valuable for understanding the influence of the response types on the interpretation of this causal parameter. This section offers an alternative representation of the CC-ACR.\footnote{In a similar fashion, the TSLS estimand has an alternative representation, as shown in Appendix \ref{app:proof_tsls_alternative}.} The representation presented here shifts the focus to changes in the treatment status and is expressed in terms of an average of responses along the causal response function. The causal response function is given by the sequence $Y^j-Y^{j-1}$ for every unit.\footnote{As mentioned by \citet{angrist1995two}, with $J+1$ treatment levels, the $\frac{(J+1) J}{2}$ potential treatment effects can be written with respect to the $J$ linearly independent treatment effects when the treatment level increases with one unit: $Y^j-Y^{j-1}$. Thus, if $D \in \{0,1,2 \} $, then $J+1=3$, meaning that there are six possible treatment effects: $Y^1-Y^0$, $Y^2-Y^0$, $Y^2-Y^1$, $Y^0-Y^1$, $Y^0-Y^2$, and $Y^1-Y^2$.} 
While the representation of the CC-ACR in this section is equivalent in terms of its definition, it offers a different perspective on the interpretation.\footnote{The representation in Theorem 1 resembles the result by \citet{frolich2007nonparametric}, whereas the representation in this section more closely resembles the result by \citet{angrist1995two}.} Of particular importance, the representation of the CC-ACR in this section introduces a weighting function that offers two advantages over the weighting function of the CC-ACR of Theorem 1: First, the weights can be estimated, and second, they also serve as a means for conducting validity tests on the LiM assumption, as will be demonstrated later in Section \ref{sec:testing_lim}. 

\newpage

\bigskip
\noindent \textbf{Proposition 2: Alternative representation of Theorem 1 \\} 
\textit{Suppose that the same conditions hold as in Theorem 1. Then,}
\begin{equation}\label{eq:thm_1_alternative}
    \begin{aligned}
        \beta_{\text{CC-ACR}} &\equiv \frac{E(Y|Z_1=Z_2=...=Z_K=1) - E(Y|Z_1=Z_2=...=Z_K=0)}{E(D|Z_1=Z_2=...=Z_K=1) - E(D|Z_1=Z_2=...=Z_K=0)} \\
        &= \sum_{j=1}^J \frac{P(D^{1...1...1} \geq j > D^{0...0...0})}{\sum_{i=1}^J P(D^{1...1...1} \geq i > D^{0...0...0})} \cdot E(Y^j - Y^{j-1} | D^{1...1...1} \geq j > D^{0...0...0}).
    \end{aligned}
\end{equation}
\bigskip

\noindent \textbf{Proof} in Appendix \ref{app:proof_theorem1_alternative}.\footnote{The proof relies on $D$ consisting of integer values in $\{ 0,1,...,J\}$, and can in some settings be obtained by a linear transformation of $D$, which boils down to multiplying the CC-ACR with a constant.} \bigskip

The weights lie between zero and one, and collectively sum up to one. 
The weights depend on the relative strength of the instruments, which is closely tied to the corresponding proportions of combined compliers. Compliers whose treatment level is moved along multiple treatment levels by the instrument contribute multiple times in the weights, which is equivalent to the representation in Theorem 1. The connection between the two representations can be seen as follows: $P(D^{1...1...1} \geq j > D^{0...0...0})=\sum_{k<j\leq l} P(T=cc_{k,l})$.

To illustrate the general result of the alternative representation of Theorem 1, consider the example of a three-valued treatment, $D \in \{0,1,2 \}$, and two binary instruments, $Z_1 \in \{ 0,1 \}$ and $Z_2 \in \{ 0,1 \}$.
Note that the combined compliers denoted by $cc_{2,0}$ increase their treatment level from zero to two when all instrument values change from zero to one, and hence these compliers contribute twice to the alternative representation in Equation (\ref{eq:thm_1_alternative}). This can be seen from the weights: These particular compliers contribute to both $P(D_i^{11} \geq 2 > D_i^{00})$ and $P(D_i^{11} \geq 1 > D_i^{00})$. 
Conversely, compliers that respond with a one-level change in the treatment due to this change in instrument values contribute only once.
The compliers in the aggregated complier group $cc_{2,1}$ contribute to $P(D_i^{11} \geq 2 > D_i^{00})$, while the compliers in the aggregated complier group $cc_{1,0}$ contribute to $P(D_i^{11} \geq 1 > D_i^{00})$. The usefulness of the weighting function in Equation (\ref{eq:thm_1_alternative}) is detailed further in the next section. 

\bigskip

\subsection{Weighting function of the alternative representation}
\label{sec:weighting}

The alternative formulation of Theorem 1, as presented in Equation (\ref{eq:thm_1_alternative}) in the previous section, offers two substantial advantages. The first advantage is that it allows for a better understanding of the estimates as it permits the estimation of the weights in Equation (\ref{eq:thm_1_alternative}), since
\begin{equation}\label{eq:weights}
    \begin{aligned}
        & P(D^{1...1...1} \geq j > D^{0...0...0}) \\
        &= P(D^{1...1...1} \geq j) -P(D^{0...0...0} \geq j) \\
        &= P(D^{0...0...0} < j) -P(D^{1...1...1} < j) \\
        &= P(D < j | Z_1=Z_2=...=Z_K=0) - P(D < j| Z_1=Z_2=...=Z_K=1) \text{\footnotemark},
    \end{aligned}
\end{equation}\footnotetext{\citeauthor{angrist1995two} (\citeyear{angrist1995two}) provide the weights for the setting with a single binary instrument.} 
\noindent where $P(D^{0...0...0} < j) -P(D^{1...1...1} < j) = P(D < j | Z_1=Z_2=...=Z_K=0) -P(D < j| Z_1=Z_2=...=Z_K=1)$ holds because of independence.

It is important to note that the weights given by the group type shares $P(T=cc_{k,l})$ of the CC-ACR estimand in Equation (\ref{eq:thm_1}) are not point-identified, meaning that these weights cannot be estimated without imposing additional assumptions. Consider, for example, the simplest case where $P(D^{11}\geq 1 > D^{00})=P(T=cc_{0,1})+P(T=cc_{0,2})$ and $P(D^{11}\geq 2 > D^{00})=P(T=cc_{1,2})+P(T=cc_{0,2})$. These are two equations with three unknowns. Ruling out one type would allow for point-identification and estimation of the shares.

The second advantage of deriving the weights in Equation (\ref{eq:thm_1_alternative}) is that necessary conditions for the validity of the LiM assumption arise from the weighting function. 
If LiM (i.e., $P(D^{1...1...1}\geq D^{0...0...0}) = 1$) holds, then it must hold that $P(D^{1...1...1} \geq j)-P(D^{0...0...0} \geq j) \geq 0$ for all $j$. Consequently, the expression in Equation (\ref{eq:weights}) must be greater than or equal to zero under LiM, meaning that the LiM assumption implies that the weighting function is positive across all treatment levels and vice versa.

\bigskip

\subsection{Identification of the CC-ACR for a continuous treatment}
\label{sec:continuous_treatment}

Up to this point, the present study has focused on discrete, ordered treatments. However, it is worth noting that in many settings the treatment can be continuous. Theorem 2 provides the results for the continuous treatment setting.

\bigskip

\noindent \textbf{Theorem 2: Continuous treatment effect}  \\
\textit{Let Assumptions 1, 2, 3, and 4 hold. If the treatment is continuous, the CC-ACR is identified as}
\begin{align*}
    & \beta_{\text{CC-ACR}} = \frac{E(Y|Z_1=Z_2=...=Z_K=1) - E(Y|Z_1=Z_2=...=Z_K=0)}{E(D|Z_1=Z_2=...=Z_K=1) - E(D|Z_1=Z_2=...=Z_K=0)}\\
    & = \int_{0}^{\infty} \frac{P(D^{1...1...1} \geq t > D^{0...0...0})}{\int_{0}^{\infty} P(D^{1...1...1} \geq j > D^{0...0...0})  dj} \cdot  \frac{\partial  E(Y^{t}| D^{1...1...1} \geq t > D^{0...0...0})}{\partial t} dt.
\end{align*}
\bigskip

\noindent \textbf{Proof} in Appendix \ref{app:proof_continuous_treatment}.
\bigskip

Theorem 2 is analogous to the CC-ACR derived for discrete, ordered treatments in Equation (\ref{eq:thm_1_alternative}). It essentially represents a weighted average derivative, with the weighting terms being determined by the shifts among combined compliers resulting from simultaneously moving all instrument values from zero to one.
The included combined complier types are those types whose treatment status $t$ lies between $D^{1...1...1}$ and $D^{0...0...0}$. Hence, the stronger the instrument, the larger the subpopulation considered by the CC-ACR. 
The weight assigned to the specific potential treatment levels is proportional to the share of complier types whose treatment status $t$ lies between $D^{1...1...1}$ and $D^{0...0...0}$.

\bigskip

\subsection{Binarizing continuous instruments}

In some applications, instruments may be continuous and can be binarized to apply the proposed methodology. The choice of cut-off value is critical, as it significantly affects the interpretation of the CC-ACR estimand by altering the complier population.

In certain cases, there is a clear rationale for selecting a cut-off, particularly in economic analyses focused on the average causal effect within a specific complier group. For example, when evaluating the impact of a new college, policymakers might set a proximity threshold, coding distances under ten minutes as zero and those over sixty minutes as one. This approach helps isolate the LATE for individuals responsive to this threshold, aiding targeted policy development.

Without such a theoretical basis, one might aim to estimate the causal effect for the largest feasible complier group. Selecting observations with extreme propensity scores captures substantial shifts in treatment assignment but reduces the number of observations. Conversely, using the median as a cut-off retains more observations but captures a smaller portion of the complier group. This trade-off requires careful consideration and a balanced approach is advisable. In the absence of economic guidance, setting the cut-off at the 25th and 75th percentiles might offer a practical compromise, likely preserving more observations while capturing a larger complier group than a median-based cut-off.

\bigskip

\section{Test for detecting violations of LiM}

\label{sec:testing_lim}

In this section, I show how the necessary conditions implied by LiM, derived in Section \ref{sec:weighting}, can be exploited to construct formal statistical tests for detecting violations of the LiM assumption. 

\bigskip

\subsection{Global violations}
\label{sec:testing_lim_global}

As demonstrated in Section \ref{sec:weighting}, under LiM, it must hold that the CDF of $D$ given $Z_1=Z_2=...=Z_K=1$ and the CDF of $D$ given $Z_1=Z_2=...=Z_K=0$ do not cross, and the former CDF first-order stochastically dominates the latter CDF. This is a necessary (though not sufficient) condition that can be verified from the data.\footnote{Note that, in a similar fashion, one can consider the choice restrictions imposed by PM and verify that these hold across the treatment margins by comparing conditional CDFs. While LiM can be tested with only one comparison, PM involves $K\cdot 2^{K-1} = 12$ comparisons of CDFs when $K=3$ instruments are available.} 
Global violations can be quickly detected through visual inspection: if the CDFs do not intersect, the necessary condition for LiM holds across all instances of the causal response function. A more formal testing procedure for stochastic dominance can be obtained through the Kolmogorov-Smirnov test, or through a multiplier bootstrap test, which is particularly of interest when the asymptotic distribution of the test statistic under the null hypothesis is unknown (see, for example, \citeauthor{abadie2002bootstrap} (\citeyear{abadie2002bootstrap})). A formal global test lies outside the scope of this paper. Instead, the next section focuses on detecting local violations.

\bigskip

\subsection{Local violations}
\label{sec:testing_lim_local}

This section addresses the possibility that more severe local violations of LiM could exist within specific subgroups and get averaged out in the full sample, which would decrease power of a test for LiM. 
It can be shown that LiM implies that the following inequality must be satisfied at any point $x$ in the covariate space (see Appendix \ref{app:inequalities_lim}):
\begin{equation}\label{eq:conditions_testing}
    E(I(D < j) | \widetilde{Z}=0, X=x) - E(I(D < j)|\widetilde{Z}=1, X=x) \geq 0 \text{ for all } j \in \{0,1,...,J \},
\end{equation}
where $\widetilde{Z}$ equals one if $Z_1=Z_2=...=Z_K=1$ and zero otherwise. Testing a condition at every level of the treatment, $j$, can offer the advantage of detecting for which causal response, $E(Y^j - Y^{j-1} | D^{1...1...1} \geq j > D^{0...0...0})$, the weight of $\tau_j$ in Equation (\ref{eq:thm_1_alternative}) might be negative. This provides knowledge at what point of the causal response function the assumption might be violated. A disadvantage is that it can be computationally intensive, especially if the treatment can attain a large range of values.

The necessary conditions established in Expression (\ref{eq:conditions_testing}) boil down to estimating heterogeneous causal effects of the instrument on the treatment.
Let $(D_i, Z_i, X_i)$ be i.i.d. observations for $i=1,...,n$, and define the pseudo variable $Q_{j,i} \equiv I(D_i<j)$.
Then, write the conditional average treatment effect (CATE) of $\widetilde{Z}$ on $Q_{j}$ at the point $X=x$ as
\begin{equation}\label{eq:cf_tau}
    \tau_{j}(x)=E(Q_{j,i}|\widetilde{Z}=1,X_i=x) - E(Q_{j,i}|\widetilde{Z}=0,X_i=x).
\end{equation} 
Under Assumptions 1 to 4, the inequality $\tau_{j}(x) \leq 0$ has to be true for every combination of $j$ and $x$. 
If $\tau_j(x)$ is positive, this indicates a violation of LiM, meaning that the necessary conditions for LiM can be interpreted as learning the sign of a conditional average treatment effect (CATE).
Moving forward, I closely follow the procedure proposed by \citet{farbmacher2022instrument} (see Appendix \ref{app:cftest_procedure} for a detailed description). 
First, a causal forest \citep{wager2018estimation} is employed for estimating these heterogeneous CATEs. Then, the heterogeneity is summarized by shallow Breiman trees, which additionally allow for visualization of the test. Relevant subgroups are selected through pruning of these trees. Finally, promising subgroups with potentially positive CATEs are selected and tests with Bonferroni-corrected p-values are performed.\footnote{In some applications, the Bonferroni correction might be too conservative and exhibit low power. This correction is most effective when tests are independent, which is clearly not the case here. Consequently, it might be more beneficial to calculate the critical value while considering the correlation between variables and tests. \citet{chernozhukov2023high} propose a bootstrap approach that allows to test on uniformly valid confidence intervals. 
In an effort to increase power, \citet{huber2022testing} implement a multiplier bootstrap which involves a score function with dimensions equal to the number of leaves tested.}

The proposed approach offers two main benefits. 
Firstly, if the degree of violation of the assumption differs for different subgroups, then this test has larger power than alternative tests, since it checks for violations of monotonicity in a specific area of the covariate space instead of in the full sample. 
Secondly, it is beneficial to form subgroups in a data-driven way, instead of having a researcher create potentially arbitrary subgroups. The latter can be especially inefficient in case of high-dimensionality of the covariate space. 

A violation of LiM can have substantial consequences, potentially leading to estimating the wrong sign of the CC-ACR parameter or to less precise estimates. These issues are particularly pronounced in the case of few compliers \citep{angrist1996identification}. LiM does allow for response types that defy with respect to some of the instruments, as long as they can be pushed towards compliance by some other instrument. However, defier types that respond most strongly to the instrument that they defy are problematic. The presence of these defiers exacerbates this bias, which is influenced by the instrument's strength and the variability of treatment effects. 
Insights into the magnitude of the violation can be gained through sensitivity tests.\footnote{For instance, \citet{klein2010heterogeneous} offers insights into recovering the LATE when monotonicity violations occur randomly and how to approximate the bias. \citet{noack2021sensitivity} develops methods to assess the sensitivity of LATEs to violations of IAM. Extending these findings to LiM violations presents an intriguing avenue for future research.} 
Detecting a violation within any subpopulation undermines the IV validity for the whole population, as it raises the potential for similar issues to exist in other subpopulations \citep{farbmacher2022instrument}.

\bigskip

\section{Estimation of the CC-ACR}

\label{sec:estimation}

\subsection{Estimation without covariates}

There are several ways of estimating the CC-ACR presented in Equation (\ref{eq:thm_1}). A simple approach is to implement the TSLS method within the subsample of observations at the outer support of the instrument values using $\widetilde{Z}=Z_1=Z_2=...=Z_K$ as the single instrument. Assuming independent sampling, this strategy yields consistent estimates and asymptotically valid confidence intervals for the parameter $\beta_\text{CC-ACR}$. 
An alternative estimation approach involves formulating moment equations and subsequently utilizing the generalized method of moments (GMM) framework. 
Furthermore, it is worth noting that the parameter in Theorem 1 can also be estimated by simply replacing the expectations with sample averages. Specifically, this involves comparing the average outcome $Y$ and the average treatment $D$ for the instrument values $\widetilde{Z}=1$ to the average outcome $Y$ and the average treatment $D$ for the instrument values $\widetilde{Z}=0$, respectively.

It is important to point out that, while increasing the number of instruments decreases the sample size used for estimating the CC-ACR, adding instruments does not inherently increase variance. This is because increasing the number of instruments can increase the share of combined compliers considered, possibly resulting in a variance reduction. For a more detailed discussion, refer to \citet{lewbel2022limited}.

Another noteworthy observation is that the proposed methodology accommodates weak instruments, provided they do not substantially reduce the number of observations and at least one strong instrument is present. As a result, valuable information from weak instruments, such as the 2-year college instrument in \citeauthor{card1995geographic}'s \citeyear{card1995geographic} study, can still be obtained.

\bigskip

\subsection{Estimation with covariates}
\label{sec:estimation_with_covariates}


Corollary 2 in Section \ref{sec:identification_covariates} provides the identification result of the CC-ACR under the assumption that conditional independence holds, assuming adequate overlap. 
While it may be tempting to incorporate covariates linearly into the subsample-TSLS approach outlined in the previous section, linear inclusion of the covariates can neglect treatment effect heterogeneity and introduce interpretation complexities already in the context of binary treatments \citep{blandhol2022tsls,sloczynski2020should}. \citet{blandhol2022tsls} demonstrate that the linear inclusion of covariates in the TSLS estimator may result in the inclusion of response types beyond compliers. It is particularly concerning that treatment effects for these additional response types might always receive negative weights.

A fully saturated first stage is necessary to preserve the interpretation of the estimates, which suggests the use of nonparametric estimation methods. With only a few discrete covariates, nonparametric approaches such as kernel regression, local polynomial regression, and series estimators may be suitable \citep{frolich2007nonparametric}. However, these methods quickly run into the curse of high dimensionality as the number of covariates increases.

To circumvent this issue, I flexibly control for covariates with rich interactions and complex functions using machine learning to handle high-dimensional controls. Specifically, I extend the double debiased machine learning (DML) framework by \citet{chernozhukov2018double} to estimate the CC-ACR with covariates of Equation (\ref{eq:cc_acr_covariates}), which equals the ratio of two average treatment effects (ATEs), in an interactive IV model.\footnote{For an introduction to DML, see \citet{chernozhukov2024applied}.}
To the best of my knowledge, this is the first paper to use the DML framework in a setting with disrete, ordered or continuous treatments and multiple binary instruments. By focusing on the subsample of size $n_s$ where either $Z_1=Z_2=...=Z_n=1$ or $Z_1=Z_2=...=Z_n=0$, and introducing the single binary instrument $\widetilde{Z}$, I reduce the estimation complexity of multiple instruments to a single instrument scenario while maintaining the advantageous interpretation of using multiple instruments. 
Furthermore, the results do not depend on binary $D$; only the instrument $\widetilde{Z}$ needs to be binary.

The following interactive instrumental variable model is considered:
    \begin{align*}
        & Y = \mu_0(\widetilde{Z},X) + \nu, \ \ \ E(\nu|\widetilde{Z},X)=0, \\
        & D = m_0(\widetilde{Z},X) + U, \ \ \ E(U|\widetilde{Z},X)=0, \\
        & \widetilde{Z} = p_0(X) + V, \ \ \ \ \ \ E(V|X)=0,
    \end{align*}
where $\nu$, $U$, and $V$ are independent. Define the following functions for the true value of the nuisance parameter $\eta_0=(\mu_0, m_0, p_0)$:
\begin{align*}
    \mu_0(\widetilde{Z},X) &= E(Y|\widetilde{Z},X), \\
    m_0(\widetilde{Z},X) &= E(D|\widetilde{Z},X), \\
    p_0(X) &= E(\widetilde{Z}|X). 
\end{align*}
The nuisance parameter $\eta=(\mu, m, p)$ denotes square-integrable functions $\mu$, $m$, and $p$. While $\mu$ and $m$ map the support of $(\widetilde{Z},X)$ to $\mathbb{R}$, $p$ maps the support of $X$ to $(\varepsilon, 1-\varepsilon)$ for some $\varepsilon\in(0,1/2)$.

Adjusting the methodology of \citet{chernozhukov2018double}, denote the true parameter of interest $\beta_0^{\text{CC-ACR}}$.
The orthogonal score for estimating $\beta_0^{\text{CC-ACR}}$ is given by
\begin{align}\label{eq:score}
    \centering
    & \psi\left((Y,D,X,\widetilde{Z});\beta^{\text{CC-ACR}},\eta\right) \nonumber \\  & \equiv \mu(1,X) - \mu(0,X) + \left( \frac{\widetilde{Z}}{p(X)} - \frac{(1-\widetilde{Z})}{1-p(X)} \right) \cdot (Y-\mu(\widetilde{Z},X)) \\ \nonumber
    & \ \ \ - \left( m(1,X) - m(0,X) + \left( \frac{\widetilde{Z}}{p(X)} - \frac{(1-\widetilde{Z})}{1-p(X)} \right) \cdot (D - m(\widetilde{Z},X)) \right) \times \beta^{\text{CC-ACR}},
\end{align}
and at $\eta_0 = (\mu_0,m_0,p_0)$ satisfies the moment condition $E(\psi((Y,D,X,\widetilde{Z});\beta_0^{\text{CC-ACR}},\eta_0)) = 0$ and the Neyman orthogonality condition $\partial_{\eta} E_{\psi}((Y,D,X,\widetilde{Z});\beta_0^{\text{CC-ACR}},\eta_0)=0$. 
Then, under similar regularity assumptions as stated in \citet{chernozhukov2018double}, the behavior of $\hat{\beta}^{\text{CC-ACR}}$ is not affected by the estimation error of the nuisance parameters:
\begin{align*}
    \sqrt{n_s}(\hat{\beta}^{\text{CC-ACR}}-\beta_0^{\text{CC-ACR}})\approx \sqrt{n_s} \mathbb{E}_n(\phi_0(Y,D,X,\widetilde{Z})),
\end{align*}
where
\begin{align*}
    \phi_0(Y,D,X,\widetilde{Z}) = -J_0^{-1} \phi(W;\beta_0^{\text{CC-ACR}},\eta_0),
\end{align*}
is the influence function and
\begin{align*}
    J_0 := E \left( m_0(1,X) - m_0(0,X) \right),
\end{align*}
the Jacobian matrix.
Subsequent these conditions, $\hat{\beta}^{\text{CC-ACR}}$ is approximately normal: 
\begin{align*}
    \sqrt{n_s} (\hat{\beta}^{\text{CC-ACR}}-\beta_0^{\text{CC-ACR}})\overset{a}{\sim} N(0,V), \ \ \ V := E(\phi_0(Y,D,X,\widetilde{Z}) \phi_0(Y,D,X,\widetilde{Z})').
\end{align*}

Estimates of $\hat{\beta}^{\text{CC-ACR}}$ are obtained through plugging in the cross-fitted residuals, constructed by the predictions of the nuisance functions of the machine learners over $K$ folds, into Equation (\ref{eq:score}). Median estimates over different sample splits can be considered to account for variability in finite samples due to sample splitting, making the estimates more robust to outliers.

Post-regularized inference proceeds as proposed by \citet{chernozhukov2018double}, with $\hat{V}$ obtained by plugging in the different components and subsequently $\sqrt{\hat{V}/n_s}$ the estimator of the standard error of $\hat{\beta}^{\text{CC-ACR}}$.

\bigskip

\section{Empirical applications}
\label{sec:results}

The proposed methodology is broadly applicable to various studies involving nonbinary, ordered treatments and multiple instruments, including studies like \cite{attanasio2013community}. My results further extend to studies using Mendelian randomization, which employs genetic variants as instruments and has found applications in social sciences and economics (see, for example, \cite{dixon2020mendelian, scholder2013child}). To provide another example, my results are relevant to applications that instrument for child BMI using the BMI of biological relatives (see, for example, \cite{cawley2004impact, lindeboom2010assessing, kline2008wages}).

To demonstrate the results of the preceding sections, I consider two empirical applications. First, I revisit \citeauthor{card1995geographic}'s (\citeyear{card1995geographic}) seminal study on the returns to education. Second, I apply the novel methodology to \citeauthor{angrist1996children}'s (\citeyear{angrist1996children}) data to study the effect of an additional child on female labor market outcomes. 

\bigskip

\subsection{The causal effect of schooling on wage}
\label{sec:results_card}

\subsubsection{Data}

Here, I briefly introduce the most relevant aspects regarding the data set, referring the interested reader to \citet{card1995geographic} for a more detailed discussion. \citet{card1995geographic} uses data from the 1979 National Longitudinal Survey of Youth (NLSY79). The data are available in the R package \textit{ivmodel} from \citet{kang2021ivmodel}. The outcome variable is the logarithm of wage, the treatment is years of education ($D \in \{ 8,9,\dots,18 \}$), and the instruments are indicators for the presence of 2-year and 4-year colleges within a county ($Z_1,Z_2 \in \{ 0,1 \}$).
Given the concerns about instrument validity raised in \citet{card1995geographic}, I adopt a similar approach by incorporating a comprehensive set of controls in the analysis. These controls include dummy variables for race, living in the South, residing in a metropolitan area, and region fixed effects to account for potential systematic variations across different geographical regions. Additionally, IQ is included as a control variable.

\citet{card1995geographic} primarily focuses on the 4-year college instrument, deeming the 2-year college instrument weak. However, excluding the 2-year college means losing information on men who comply only with this instrument. My methodology can handle weak instruments as long as one is strong. While including the 2-year college reduces observations for CC-ACR estimation by 44\% (see Table \ref{tab:num_obs}), it adds compliers, potentially offsetting the loss of observations without reducing efficiency (see \citet{lewbel2022limited}).

\begin{table}[btp]
    \centering
    \caption{Number of observations in the full sample and the subsample used for estimating the CC-ACR in the study by \citeauthor{card1995geographic} (\citeyear{card1995geographic}).}
    \label{tab:num_obs}
    \begin{tabular}{lcc} \toprule
          & Nr. obs. & \% obs. \\ \midrule
         Full sample & 2,061 & 100\% \\
         Subsample where $\widetilde{Z}$ equals zero or one & 1,159 & 56\% \\ \bottomrule
    \end{tabular}
\end{table}

\bigskip

\subsubsection{Analysis of the causal effect of schooling on wage}
\label{sec:estimation_results}

\begin{figure}
    \centering
    \includegraphics[width=0.825\linewidth]{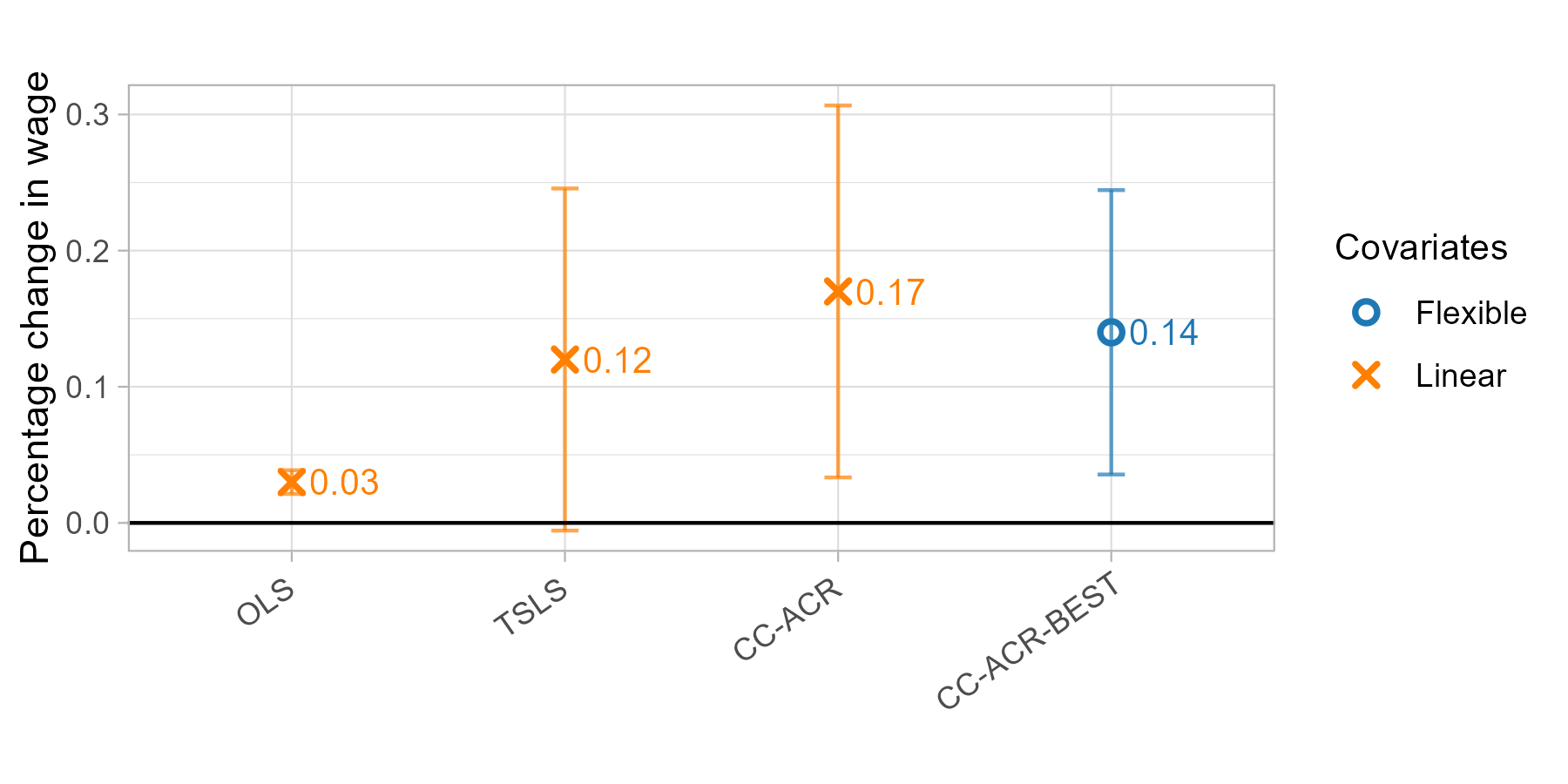}
    \caption{This figure presents some of the estimates of the causal effect of schooling on wage reported in Table \ref{tab:all_estimates_schooling} for easier comparison. CC-ACR-BEST is the DML estimate obtained using the machine learner that achieved the lowest sum of standardized mean RMSE}
    \label{fig:main_estimates_card}
\end{figure}

In this section, I present the empirical findings from implementing the TSLS methodology as well as the proposed CC-ACR methodology. 
The results are reported in Table \ref{tab:all_estimates_schooling} and Figure \ref{fig:main_estimates_card}.
The TSLS specification is saturated in the instruments in the first stage. The TSLS estimates without covariates should be interpreted as the weighted average of the weighted averages in Equation (\ref{eq:beta_tsls}) of Proposition 1. 
It is important to note that differences in estimates can be attributed to differences in underlying complier populations, underlying assumptions, potential violations of the partial monotonicity assumption, or the different weighting schemes. Notably, the confidence intervals of TSLS are comparable to those of CC-ACR, despite CC-ACR using only 56\% of the observations. However, it should be noted, that the two methods estimate different causal parameters.

Columns 3 and 4 present the results when using the 2-year and 4-year college instruments individually, while controlling for the other instrument. The estimated effects are imprecise, likely because each instrument lacks sufficient strength. However, when the estimates are combined - shown in Column 6 with the TSLS estimate and Column 8 with the CC-ACR estimate - the estimated effects are significant, leveraging the combined strength of the instruments.

Covariates are included linearly in Columns 6 and 8 of Table \ref{tab:all_estimates_schooling}. This linear inclusion may be problematic in the case of multiple instruments, as it can introduce biases from never-takers and always-takers, potentially contaminating the estimated averages of combined complier effects (see Section \ref{sec:estimation_with_covariates}).
Ideally, one would include all covariate and instrument interactions to maintain optimal interpretability. However, this fully saturated specification suffers from high dimensionality. Specifically, ignoring the control for IQ for the moment, the number of parameters in a fully saturated first stage with 2 binary instruments and 12 dummies is $2^{14} = 16,384$, which far exceeds the number of observations in the subsample, making it infeasible to use a fully saturated specification.

I now shift my focus to the DML approach discussed in Section \ref{sec:estimation_with_covariates}, which offers greater flexibility in accounting for confounding factors, thereby maintaining the interpretation of the CC-ACR. For estimating the nuisance functions, I employ three different machine learning algorithms: Lasso  \citep{tibshirani1996regression}, Random Forest \citep{breiman2001random}, and Boosted Trees \citep{friedman2001greedy}. Details regarding specific implementations, including hyperparameter tuning, can be found in Appendix \ref{app:specifications_schooling}.

The best learner was selected by standardizing the root mean squared error (RMSE) for predicting the three nuisance functions and choosing the learner that minimized the sum of these errors (see Table \ref{tab:rmse_ml_card} in Appendix \ref{app:specifications_schooling}). The estimate using the best-performing learner (Boosted Trees) is presented in Figure \ref{fig:main_estimates_card}, though the mean RMSE values across all machine learners are similar (see Table \ref{tab:rmse_ml_card} in Appendix \ref{app:specifications_schooling}), as well as the CC-ACR estimates, indicating that the primary insights are unaffected by the choice of learner.

The CC-ACR estimate with the best learner implies that a one-year increase in education is associated with an average wage increase of approximately 14\% for the combined complier population.
This is slightly higher than the estimate found with TSLS, though it should be noted that TSLS requires imposing PM instead of LiM, and PM might be violated, leading to an undesirable interpretation.
The large effect might be attributed to the inclusion of the 2-year college, which adds compliers likely at the lower end of the wage distribution. The CC-ACR estimate therefore includes individuals who might benefit more from schooling than those affected by the presence of a 4-year college or those unaffected by the presence of a college, aligning with earlier arguments theorized by \citet{card1995geographic}.
This is particularly interesting because it provides insights about the 2-year college compliers, even though the 2-year college instrument is a weak instrument and does not provide consistent estimates by itself. Adding 2-year college compliers is possible thanks to the strong 4-year college instrument, meaning that CC-ACR using both instruments offers additional insights.

Finally, the CC-ACR estimates for the years of schooling variable account for individuals who attended some college but did not complete a degree, in contrast to the binary college diploma variable, which only indicates the absence of a degree. Exploring how these differences in estimates could support either human capital theory (education enhances productivity) or signaling theory (education signals ability to employers) would be valuable but is beyond the scope of this study.

\begin{table}[!btp]
\centering
\caption{The causal effect estimates of an additional year of schooling on wage.}
\label{tab:all_estimates_schooling}
\begin{adjustbox}{max width=\textwidth}
\begin{tabular}{@{}lccccccccccc@{}}
\toprule
     & \multicolumn{2}{c}{$\hat{\beta}_{\text{OLS}}$} & $\hat{\beta}_{\text{2-year}}$ & $\hat{\beta}_{\text{4-year}}$ & \multicolumn{2}{c}{$\hat{\beta}_{\text{TSLS}}$} & \multicolumn{2}{c}{$\hat{\beta}_{\text{CC-ACR}}$} & $\hat{\beta}_{\text{DML-Lasso}}$ & $\hat{\beta}_{\text{DML-RF}}$ & $\hat{\beta}_{\text{DML-Boosted}}$ \\ 
     & (1) & (2) & (3) & (4) & (5) & (6) & (7) & (8) & (9) & (10) & (11) \\
      \midrule
    \textit{Years of schooling} & $0.039^{***}$ & $0.027^{***}$  &  0.286  &  0.072 &  $0.254^{***}$ &  $0.117^{*}$ & $0.268^{***}$ & $0.170^{*}$ & $0.147^{***}$ & $0.122^{***}$ & $0.136^{**}$  \\
    (Std. err.) & (0.004) & (0.004) & (0.217) & (0.069) & (0.060) & (0.064) & (0.068) & (0.070) & (0.047) & (0.042) & (0.053) \\ 
    Observations & 2,061  & 2,061  &  2,061 &  2,061 &  2,061 & 2,061  &  1,159 & 1,159 & 1,159  &  1,159 & 1,159 \\
    \% observations & 100\% & 100\% & 100\% & 100\% &  100\% &  100\% &  56\%  &  56\% &  56\% &  56\%  &  56\% \\
    Covariates & no & linear & linear & linear & no & linear & no & linear & flexible & flexible & flexible \\
     \bottomrule
\multicolumn{12}{p{1.4\textwidth}}{\footnotesize \textit{Note}. This table presents the estimates of causal effects of an additional year of schooling on wage for different causal parameters and estimation approaches. $\hat{\beta}_{\text{2-year}}$ and $\hat{\beta}_{\text{4-year}}$ give the instrument-specific LATE when using the instruments separately. For columns (9), (10), and (11), results are obtained using five-fold cross-fitting. For columns (9), (10), and (11), median estimates and standard errors across 25 splits are reported to take into account different sample splits.} \\
\multicolumn{12}{p{\textwidth}}{\footnotesize *Significance level: * $p<0.1$, ** $p<0.05$, *** $p<0.01$.}
\end{tabular}
\end{adjustbox}
\end{table}

\bigskip

\subsubsection{Weighting function of the CC-ACR}

\begin{figure}[!p]
    \centering
    \includegraphics[width=0.8\textwidth]{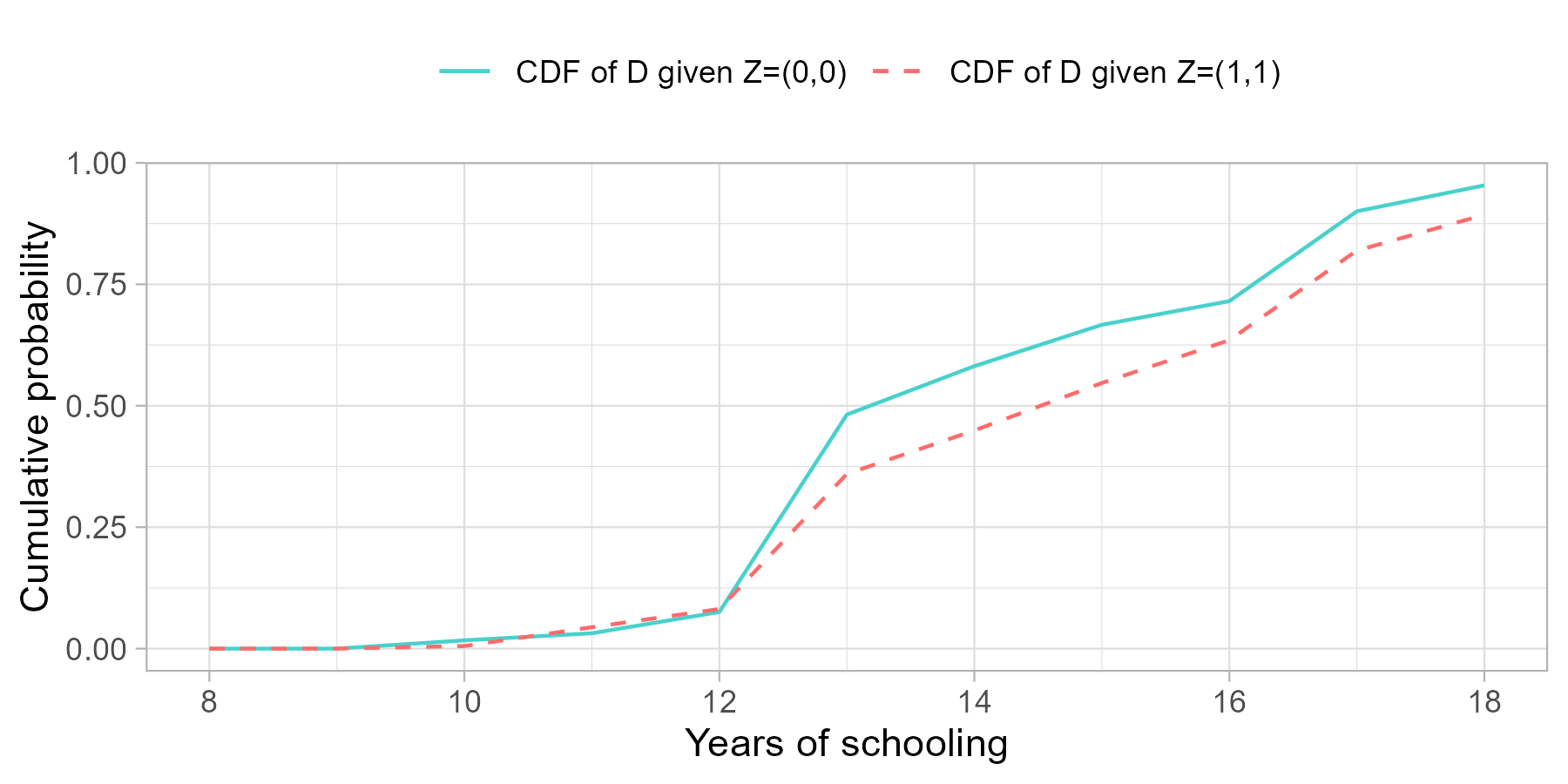}
    \caption{CDFs of the treatment conditional on the outer support of the instrument values in \citeauthor{card1995geographic}'s (\citeyear{card1995geographic}) application. 
    When the CDFs cross, this might provide a visual indication that the necessary condition for LiM does not hold at that treatment level.
    }
    \label{fig:cdf_lim_card}
\end{figure}

\begin{figure}[!bthp]
    \centering
    \includegraphics[width=0.8\textwidth]{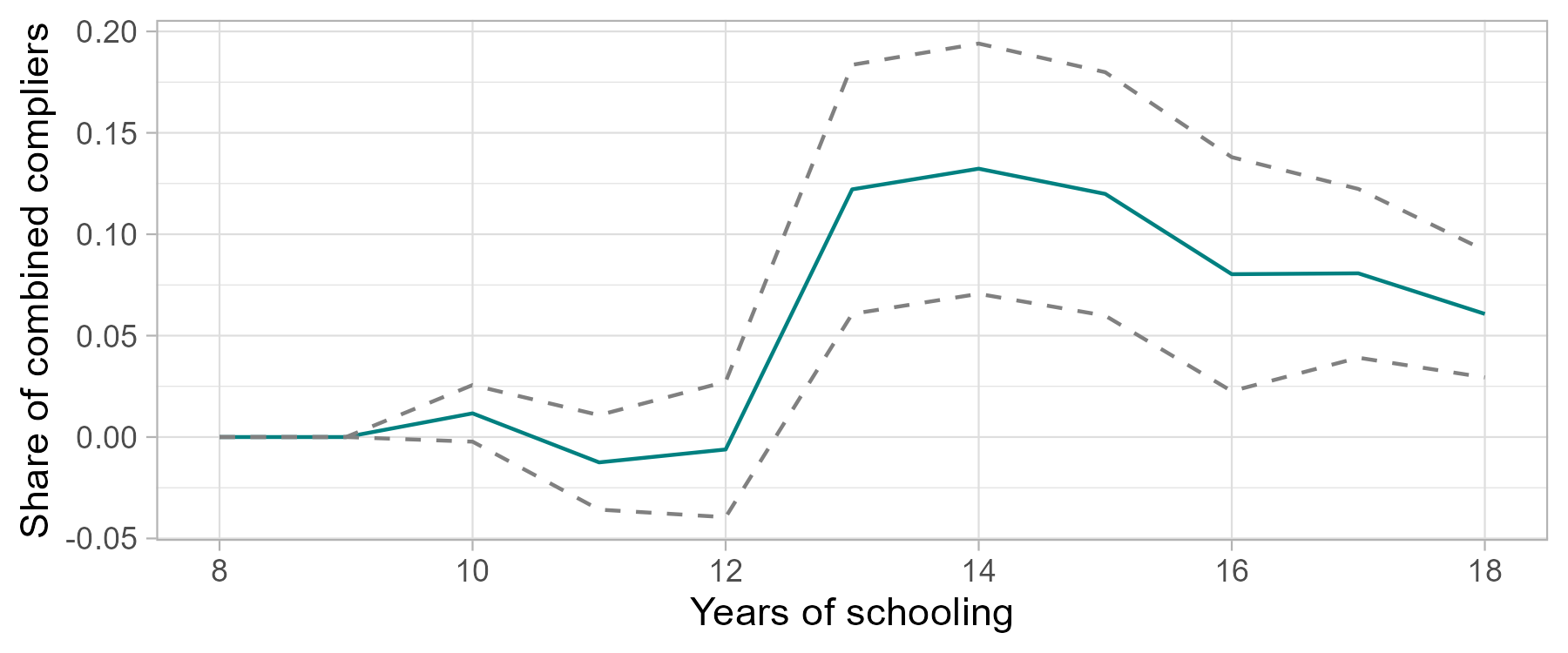}
    \caption{The unstandardized weighting function of the CC-ACR parameter of Equation (\ref{eq:thm_1_alternative}) in \citeauthor{card1995geographic}'s (\citeyear{card1995geographic}) application. 
    Standardizing these values to sum up to one gives the weights on the average causal responses of the combined complier types. 
    95\% confidence intervals are calculated in a standard fashion for a difference in proportions and indicated by the dashed lines.}
    \label{fig:weighting_card}
\end{figure}

As shown in Section \ref{sec:weighting}, the unstandardized weighting function can be obtained by subtracting the CDFs of the treatment conditional on the instrument values as follows: $P(D < j | Z_1=Z_2=...=Z_K=0)$ and $P(D < j | Z_1=Z_2=...=Z_K=1)$.
Figure \ref{fig:cdf_lim_card} depicts the two CDFs and Figure \ref{fig:weighting_card} depicts their unstandardized difference. Standardizing this to sum to one (that is, normalizing them by the first stage), this function provides the weights associated with the per-level effects along the causal response function of the CC-ACR as presented by Equation (\ref{eq:thm_1_alternative}). 
These weights reflect the combined strength of the instruments and are informative about the complier distribution across the range of treatment values. 

For each year of schooling $j$, the difference is the share of individuals whose education increases from less than $j$ years to $j$ years or more in response to the shift in instrument values. 
Unsurprisingly, the figure shows that more weight is concentrated around 12 to 14 years of education. This period, just after high school, is when most individuals make decisions about attending college, and thus, when the instruments (such as the presence of colleges) are most likely to influence educational attainment. This observation aligns with the findings of \citet{kling2001interpreting}.
While it is not possible to determine the exact size of the complier population due to potential overlaps at different treatment values, Figure \ref{fig:weighting_card} offers an estimate of a lower bound. The data indicates that at least 13\% (the maximum value of shares) belongs to the combined complier population.

\bigskip

\subsubsection{Plausibility of LiM}

In \citeauthor{card1995geographic}'s \citeyear{card1995geographic} study, there are $(J+1)^{2^K} = 11^4 = 14,641$ different initial response types (see Table \ref{tab:number_of_types}). Identifying a causal effect necessitates ruling out some types by imposing monotonicity. LiM is arguably more plausible than PM in this context.

To illustrate, consider the two types in Table \ref{tab:lim_vs_pm_schooling}. The first type, $t_{12,14,12,14}$, does not attend college when no college is nearby ($D^{00} = 12$), attends a 2-year college if one is nearby ($D^{01} = 14$), but does not respond to the presence of a 4-year college ($D^{10} = 12$). The second type, $t_{12,14,16,14}$, does not attend college when no college is nearby ($D^{00} = 12$), attends a 2-year college if one is nearby ($D^{01} = 14$), and attends a 4-year college if one is nearby ($D^{10} = 16$).
Both types are likely to exist, but PM would rule out one, as it requires either $P(D^{10} \geq D^{11}) = 1$ or $P(D^{10} \leq D^{11}) = 1$. 
While PM requires weak monotonicity in the propensity score function with respect to instruments, LiM does not, and thus allows for the coexistence of the two example types in Table \ref{tab:lim_vs_pm_schooling} and many more.

Interestingly, \citet{mogstad2021causal} highlight that a key limitation of IAM is its inherent preference for one instrument over another, which restricts choice behavior. However, when extending PM to the nonbinary treatment setting, PM faces similar restrictions.

\begin{table}[btp]
    \centering
    \caption{Number of initial response types for different combinations of treatment and instrument values.}
    \label{tab:number_of_types}
    \begin{tabular}{@{}ccc@{}}
        \toprule
        Treatment \hspace{0.5cm} & Instrument(s) \hspace{0.5cm} & Nr. Response Types \\
        \midrule
        $D \in \{0,1\}$ & $Z_1 \in \{ 0,1\}$ & 4 \\[0.5em]
        $D \in \{0,1\}$ & $Z_1, Z_2 \in \{ 0,1\}$ & 16 \\[0.5em]
        $D \in \{0,1,2\}$ & $Z_1,Z_2 \in \{ 0,1\}$ & 81 \\[0.5em]
        $D \in \{ 8,9,...,18\}$ & $Z_1,Z_2 \in \{ 0,1\}$ & 14,641 \\
        \bottomrule
    \end{tabular}
\end{table}

\begin{table}[btp]
    \centering
    \caption{Two example types in \citeauthor{card1995geographic}'s (\citeyear{card1995geographic}) study. 
    PM rules out one of these two types, imposing that individuals either prefer the 2-year college or the 4-year college. Both types are consistent with LiM ($P(D^{11}\geq D^{00})=1$). $Z_1$ is the instrument for the proximity of a 2-year college, and $Z_2$ for the proximity of a 4-year college.
    }
    \label{tab:lim_vs_pm_schooling}
    \begin{tabular}{l|cccc} \hline
        Type & $D^{00}$ & $D^{01}$ & $D^{10}$ & $D^{11}$ \\ \hline
        $t_{12,14,12,14}$ & 12 & 14 & 12 & 14 \tabularnewline
        $t_{12,14,16,14}$ & 12 & 14 & 16 & 14 \\ \hline
    \end{tabular}
\end{table}

\bigskip

\subsubsection{LiM test}
\label{sec:limtest_card}

LiM can be violated if there exist defiers with respect to one instrument who cannot be convinced to comply by another instrument. Consider the case of defiers with respect to the 4-year college instrument. These individuals may be negatively influenced by proximity to a 4-year college due to negative interactions with the college community, unfavorable perceptions of the institution's culture, or discouraging information about attending college. Furthermore, familiarity with the nearby college might diminish its perceived prestige, deterring them from pursuing a 4-year education. If such defiers cannot be persuaded to comply by the presence of a 2-year college, LiM is violated.

A visual inspection of Figure \ref{fig:cdf_lim_card} hints at a potential violation of LiM for low treatment values, as the CDFs cross when the number of years if schooling is below 12 years. To address potential violations of LiM within subgroups of observed characteristics, I implement the local LiM test as described in Section \ref{sec:testing_lim_local}. For this, I adapt the R package \textit{LATEtest} developed by \citet{farbmacher2022instrument}.\footnote{\label{note1}The procedure's configurations are as follows: The fraction of data used for each tree equals its default value of 0.5, and the minimum size of control and treated observations per leaf is set to 50.} The test indicates no detected violation of the LiM assumption. 

\bigskip

\subsection{The causal effect of additional child on female labor market outcomes}
\label{sec:results_twins}

In this section, I apply the proposed methodology to the data from \citet{angrist1996children} to analyze the impact of an additional child on female labor market outcomes.

\subsubsection{Data}

The data for the analysis is derived from the 1980 Census Public Use Micro Data Samples (PUMS).
I consider the sample of married women. For an in-depth discussion of the data, see \citet{angrist1996children}.
The primary outcome variables considered are a mother's annual labor income, the hours worked per week, and the weeks worked per year. 
The treatment variable is the number of children ($D \in \{ 2,...,6 \}$). Families with more than six children are excluded to avoid low representation when sample splitting for the DML-based estimation approach, to aid in the interpretation of the estimated CC-ACR, and since the data only contains information for the first five children born.

I consider two instruments for the analysis. The first instrument, $Z_1$, is a binary variable that equals one if there are twins at the second or subsequent births and zero otherwise. To the best of my knowledge, this instrument has not been previously used in the literature. Unlike the twinning instrument by \citet{angrist1996children}, which is specific to the birth of a third child, this novel instrument provides incentives for having an additional child for any number of children between 2 and 6. The second instrument, $Z_2$, is the classical same-sex instrument, which is equal to one when the first two children are of the same sex.\footnote{Note that using this instrument is equivalent to using an instrument that equals one if the first two, three, four, or more children are all of the same sex and zero otherwise.} 

Following \citet{angrist1996children}, the covariates included in the analysis are mother's age, age at first birth, sex of the first-born, and indicators for race and Hispanic ethnicity. Additionally, the analysis controls for the sex of the second-born, the marital status of the mothers, and the mothers' highest level of educational attainment.

\bigskip

\subsubsection{Analysis of the causal effect of an additional child on female labor market outcomes}
\label{sec:estimation_results_twins}

\begin{figure}[tbp]
    \centering
    \begin{subfigure}[b]{0.8\textwidth}
        \centering
        \includegraphics[width=\linewidth]{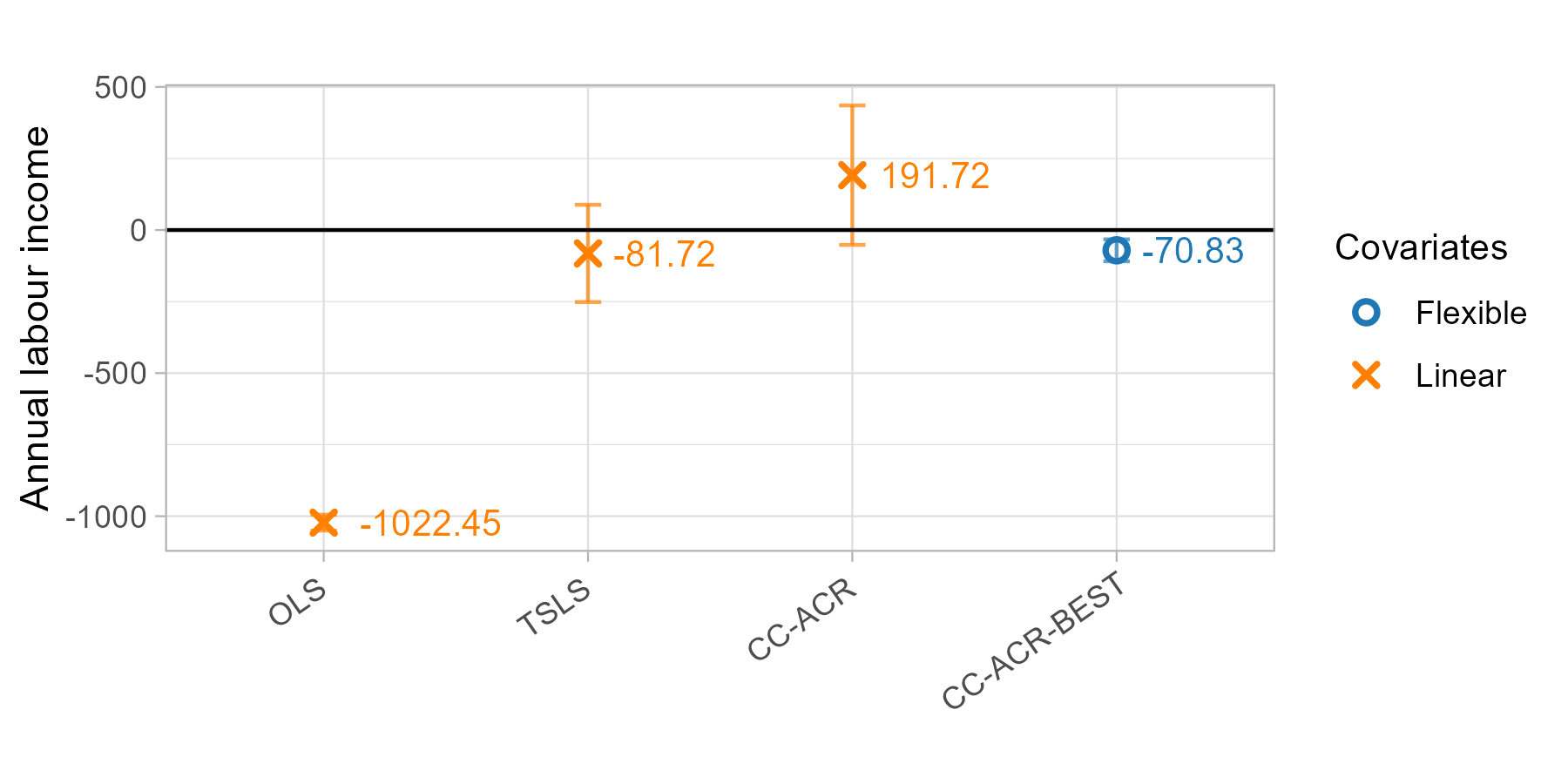}
        \caption{Causal effect of an additional child on annual labor income.}
        \label{fig:main_estimates_children_a}
    \end{subfigure}
    
    \vspace{0.75cm} 
    
    \begin{subfigure}[b]{0.8\textwidth}
        \centering
        \includegraphics[width=\linewidth]{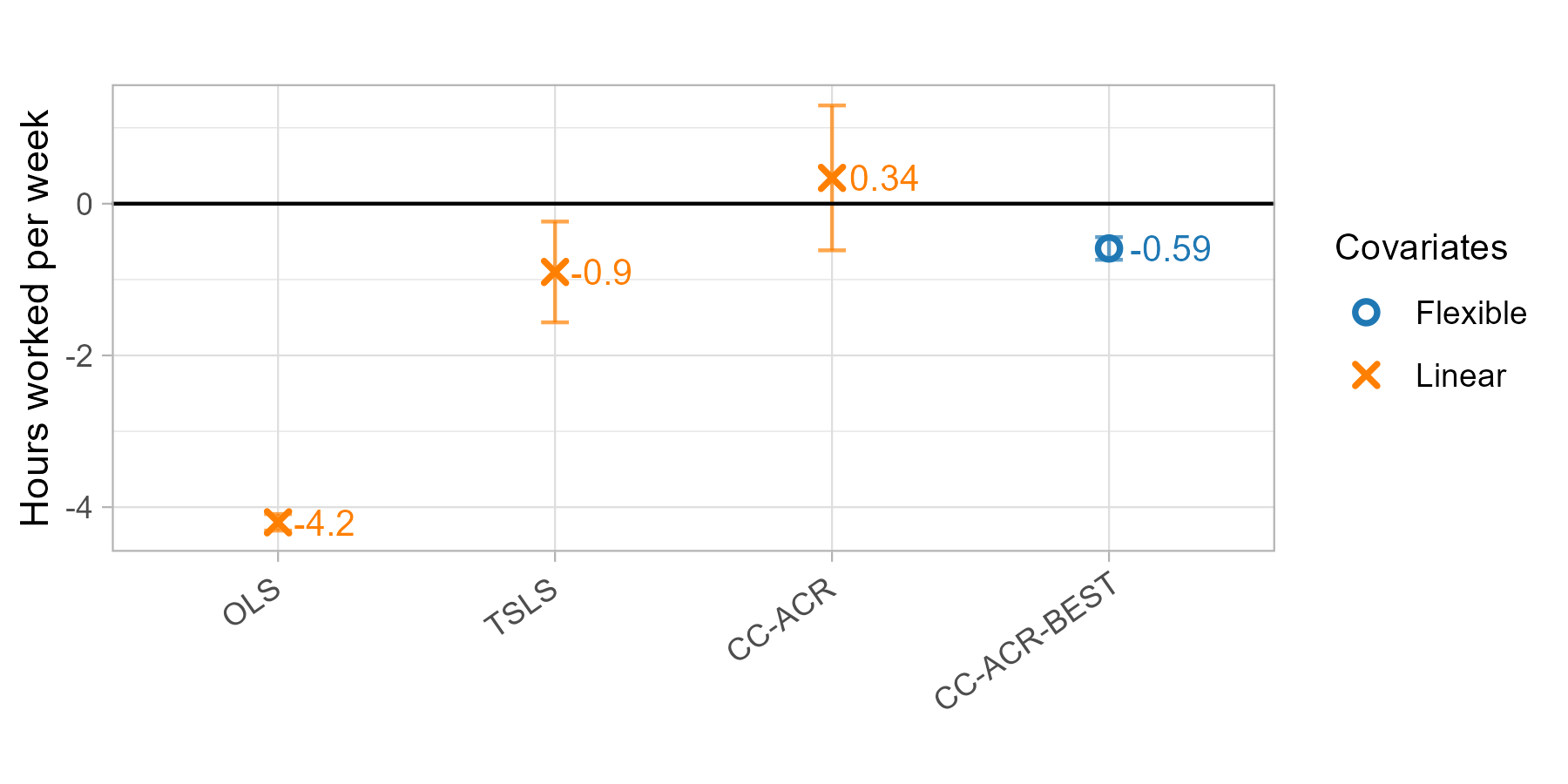}
        \caption{Causal effect of an additional child on hours worked per week.}
        \label{fig:main_estimates_children_b}
    \end{subfigure}
    
    \vspace{0.75cm} 
    
    \begin{subfigure}[b]{0.8\textwidth}
        \centering
        \includegraphics[width=\linewidth]{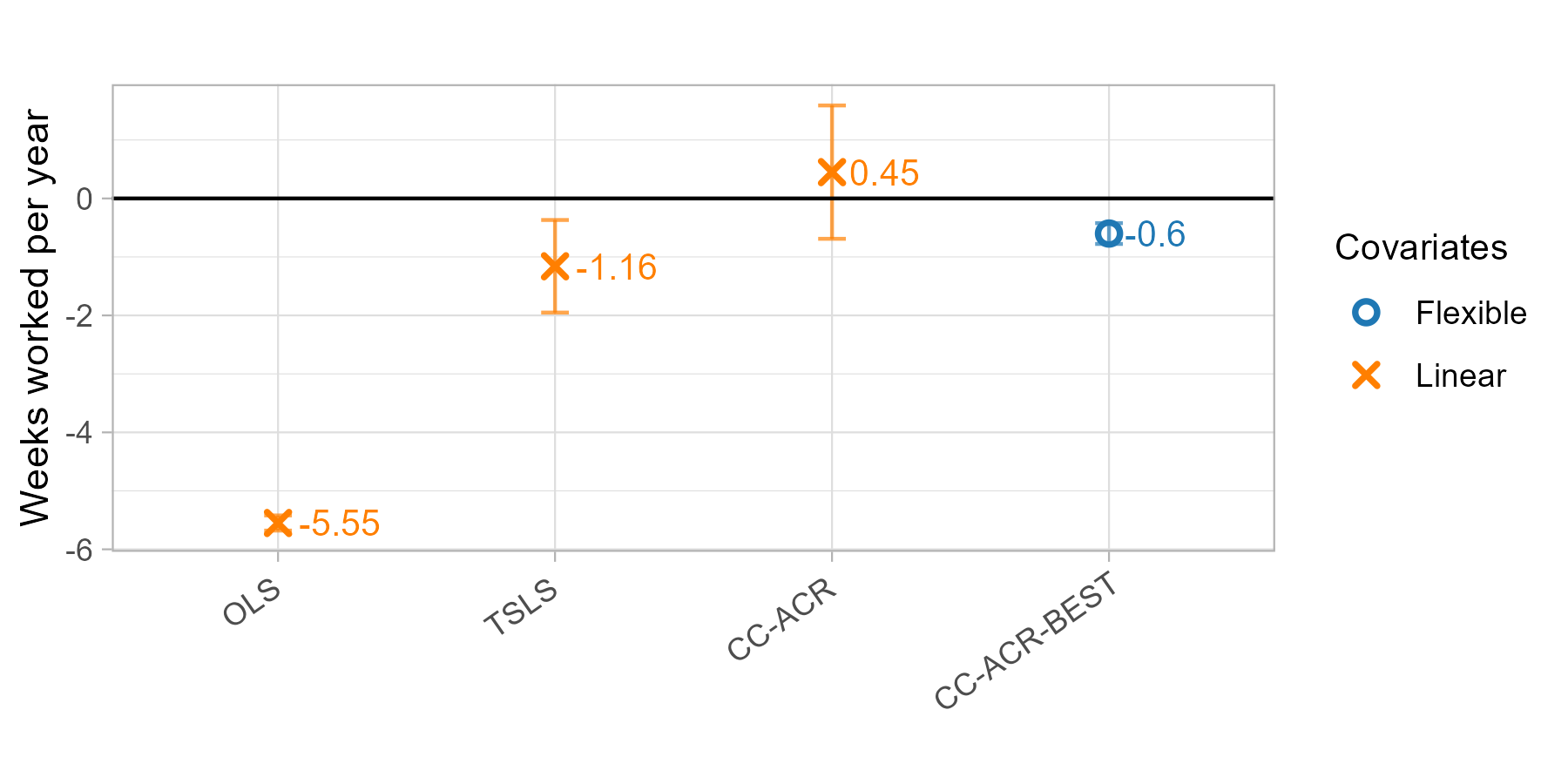}
        \caption{Causal effect of an additional child on weeks worked.}
        \label{fig:main_estimates_children_c}
    \end{subfigure}
    
    \caption{This figure presents some of the estimates reported in Table \ref{tab:all_estimates_children} for easier comparison. CC-ACR-BEST is the DML estimate obtained using the machine learner that achieved the lowest sum of standardized mean RMSE.}
    \label{fig:main_estimates_children}
\end{figure}

This section presents the causal effects of an additional child on three female labor market outcomes: annual labor income, hours worked per week, and weeks worked per year, using different estimation techniques.
Table \ref{tab:all_estimates_children} provides estimates, some of which are emphasized in corresponding Figure \ref{fig:main_estimates_children}.

When using the novel twinning instrument that considers twinning at any birth (Column 3), no significant effect is observed. This contrasts with the original paper, which focused on twinning at the second birth. One possible explanation for this difference is that women who have an additional child at, for example, the fourth birth might already be inclined to have larger families and be less active on the labor market. As a result, compliers with respect to the novel twinning instrument likely experience a smaller effect of an additional child on their labor market outcomes.

The TSLS and CC-ACR estimates combining both instruments and linearly including covariates, presented in Columns 6 and 8, respectively, indicate no effect of an additional child on annual labor income (Panel A), and only the TSLS estimate indicates a reduction of 0.896 hours worked per week.
Columns 9 and 11 display the results of the CC-ACR estimates obtained using DML, which flexibly accounts for covariates. Detailed specifications for the machine learning models are provided in Appendix \ref{app:specifications_twins}. Tree-based methods reveal a significant, albeit small, reduction in labor income between \$97.245 and \$70.832 (Panel A), with the latter estimate produced by using Boosted Trees as learners for the nuisance parameters, which achieved the minimum combined RMSE. In contrast, with Lasso as learner, the effect estimates are not significant. This may be attributed to only including third-order interactions (see Appendix \ref{app:specifications_twins}), which suggests that the poor performance of Lasso could be due to the highly complex nature of the relationships involved. Across the machine learning methods, a significant reduction in hours worked per week is observed, ranging from 0.59 to 0.895 hours (Panel B). Additionally, weeks worked decreased significantly, with reductions between 0.61 and 0.94 weeks (Panel B). Again, Boosted Trees outperform the other learners in terms of the RMSE (see Appendix \ref{app:specifications_twins}) and the effect estimates indicate a significant reduction of 0.59 hours worked per week and 0.61 weeks worked.  

\begin{table}[tbp]
\centering
\caption{The causal effect estimates of an additional child on female labor market outcomes.}
\label{tab:all_estimates_children}
\begin{adjustbox}{max width=\textwidth}
\begin{tabular}{@{}lccccccccccc@{}}
\toprule
     & \multicolumn{2}{c}{$\hat{\beta}_{\text{OLS}}$} & $\hat{\beta}_{\text{twinning}}$ & $\hat{\beta}_{\text{same-sex}}$ & \multicolumn{2}{c}{$\hat{\beta}_{\text{TSLS}}$} & \multicolumn{2}{c}{$\hat{\beta}_{\text{CC-ACR}}$} & $\hat{\beta}_{\text{DML-Lasso}}$ & $\hat{\beta}_{\text{DML-RF}}$ & $\hat{\beta}_{\text{DML-Boosted}}$ \\ 
     & (1) & (2) & (3) & (4) & (5) & (6) & (7) & (8) & (9) & (10) & (11) \\
      \midrule
      \multicolumn{12}{l}{\textbf{Panel A: Causal effect on labor income}} \\ \midrule
    \textit{Additional child} & $-709.900^{***}$ &  $-1022.451^{***}$ &  44.864  &  $-631.253^{***}$ &  -143.230 &  -81.720 & 40.500 & 191.718 & 19.016 & $-97.245^{***}$ &  $-70.832^{***}$ \\
    (Std. err.) & (13.930) & (14.203) & (93.878) & (229.681) & (95.140) & (86.734) & (14.200) & (124.279) & (20.723) & (23.309) & (19.574) \\ 
    Observations & 207,674  & 207,674  &  207,674 &  207,674 &  207,674 & 207,674  &  103,935 & 103,935 & 103,935  &  103,935 & 103,935 \\
    \% observations & 100\% & 100\% & 100\% & 100\% &  100\% &  100\% &  50\%  &  50\% &  50\% &  50\%  &  50\% \\
    Covariates & no & linear & linear & linear & no & linear & no & linear & flexible & flexible & flexible \\
     \midrule
\multicolumn{12}{l}{\textbf{Panel B: Causal effect on hours worked per week}} \\ \midrule
    \textit{Additional child} & $-2.468^{***}$ & $-4.137^{***}$ & -0.167  & $-4.055^{***}$  & -1.780  &  $-0.896^{**}$ & $-1.167^{*}$ &  0.340 & $-0.895^{***}$ & $-0.873^{***}$ & $-0.590^{***}$  \\
    (Std. err.) & (0.054) & (0.056) & (0.372) & (0.981) & (0.367) & (0.339) & (0.552) & (0.487) & (0.080) & (0.091) & (0.076) \\ 
    Observations & 207,674  & 207,674  &  207,674 &  207,674 &  207,674 & 207,674  &  103,935 & 103,935 & 103,935  &  103,935 & 103,935 \\
    \% observations & 100\% & 100\% & 100\% & 100\% &  100\% &  100\% &  50\%  &  50\% &  50\% &  50\%  &  50\% \\
    Covariates & no & linear & linear & linear & no & linear & no & linear & flexible & flexible & flexible \\
     \midrule
\multicolumn{12}{l}{\textbf{Panel C: Causal effect on weeks worked}} \\ \midrule
    \textit{Additional child} & $-3.590^{***}$ & $-5.554^{***}$ &  -0.316 & $-5.033^{***}$ &  $-2.141^{***}$ &  $-1.160^{**}$ & $-1.278^*$ & 0.452 & $-0.944^{***}$ & $-0.940^{***}$ & $-0.605^{***}$  \\
    (Std. err.) & (0.065) & (0.066) & (0.444) & (1.166) & (0.441) & (0.404) & (0.660) & (0.582) & (0.096) & (0.108) & (0.091) \\ 
    Observations & 207,674  & 207,674  &  207,674 &  207,674 &  207,674 & 207,674  &  103,935 & 103,935 & 103,935  &  103,935 & 103,935 \\
    \% observations & 100\% & 100\% & 100\% & 100\% &  100\% &  100\% &  50\%  &  50\% &  50\% &  50\%  &  50\% \\
    Covariates & no & linear & linear & linear & no & linear & no & linear & flexible & flexible & flexible \\
     \bottomrule 
\multicolumn{12}{p{1.65\textwidth}}{\normalsize \textit{Note}. This table presents the estimates of causal effects of an additional child on female labor market outcomes for different causal parameters and estimation approaches.. $\hat{\beta}_{\text{twinning}}$ and $\hat{\beta}_{\text{same-sex}}$ give the instrument-specific LATE when using the instruments separately. For columns (9), (10), and (11), results are obtained using five-fold cross-fitting. For columns (9), (10), and (11), median estimates and standard errors across 5 splits are reported to take into account different sample splits.} \\
\multicolumn{12}{p{\textwidth}}{\normalsize *Significance level: * $p<0.1$, ** $p<0.05$, *** $p<0.01$.}
\end{tabular}
\end{adjustbox}
\end{table}

The results presented in this section complement those of \citet{angrist1996children}.
The CC-ACR estimates the average effect of having any additional child after the first, which differs from the original study by \citet{angrist1996children} that focuses on the effect of having more than two children. Studies using the twinning instrument at the second birth or the same-sex of the first two children are restricted to estimating the impact of a third child. In contrast, the CC-ACR approach captures the effect of any additional child after the first, as the novel twinning instrument affects the likelihood of having a second child or more, while the same-sex instrument influences the birth of a third child or more.
Second, \citet{angrist1996children} employ each instrument individually. A key advantage of using multiple instruments, as done in this study, is that it accommodates defiers with respect to the same-sex instrument. These defiers do not affect the CC-ACR estimates as long as they can be induced to comply through the influence of the twinning instrument.

With respect to annual labor income, \citet{angrist1996children} report reductions of \$1,338 using the same-sex instrument and \$1,308 using the twinning instrument for married women, specifically for the effect of having a third child. For hours worked per week, they find reductions of 4.87 and 4.59 hours, respectively. 
For weeks worked, they find reductions of 5.45 and 5.15 weeks, respectively. 
When compared to the CC-ACR estimates in Column 11 of Table \ref{tab:all_estimates_children}, the negative effect of an additional child, when considering up to six children, is noticeably smaller than the effect of a third child as estimated by \citet{angrist1996children}. This discrepancy likely arises because women who, for instance, have twins at the fourth or fifth birth may have different labor market preferences and are already predisposed to lower labor market participation, thereby attenuating the observed effect.

\bigskip

\subsubsection{Weighting function of the CC-ACR}

\begin{figure}[!bthp]
    \centering
    \includegraphics[width=0.8\textwidth]{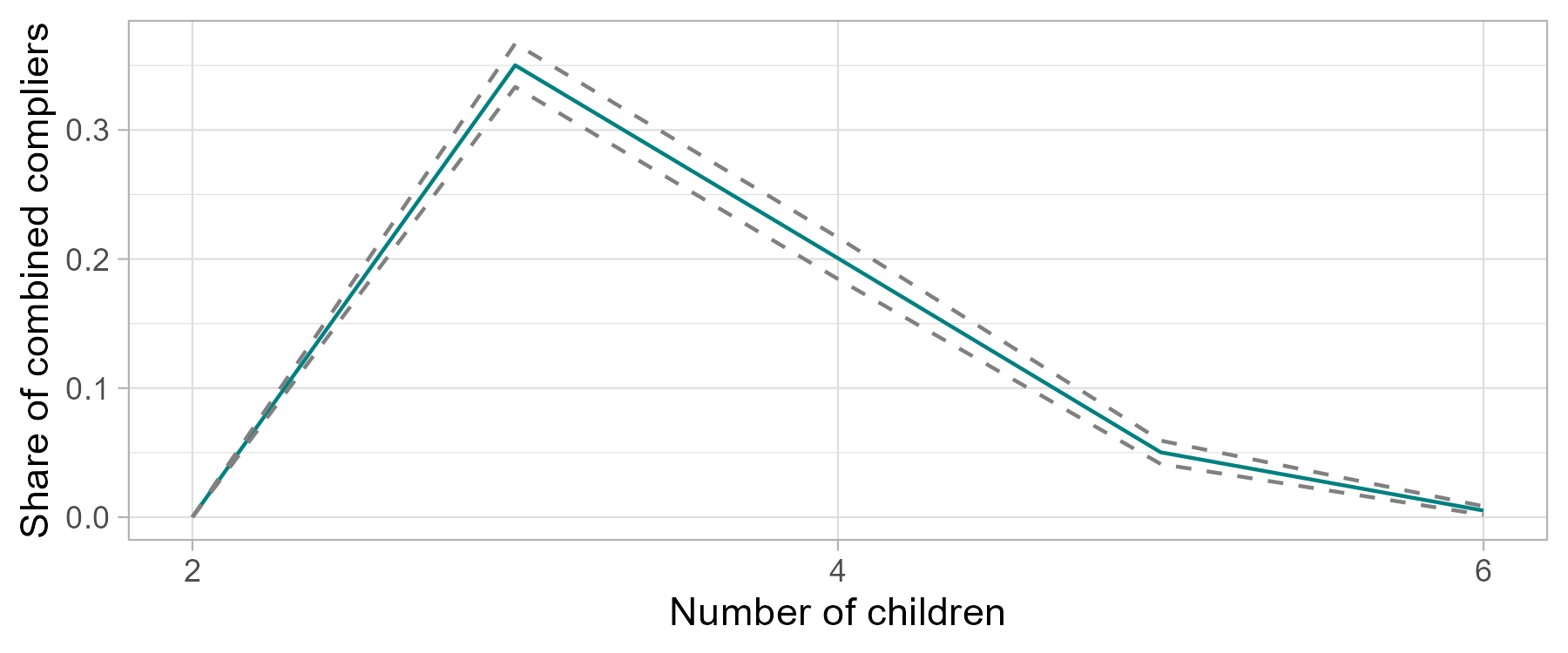}
    \caption{The unstandardized weighting function of the CC-ACR parameter of Equation (\ref{eq:thm_1_alternative}) in \citeauthor{angrist1996children}'s (\citeyear{angrist1996children}) application. 
    Standardizing these values to sum up to one gives the weights on the average causal responses of the combined complier types. 
    95\% confidence intervals are calculated in a standard fashion for a difference in proportions and indicated by the dashed lines.}
    \label{fig:weighting_twins}
\end{figure}

The CC-ACR weighting function provides insights into the shares of women at different household sizes who have more children because of the instrument incentives. Figure \ref{fig:weighting_twins} shows an absence of compliers at the treatment level of two children, which is expected as the instruments incentivize having three or more children. Additionally, approximately 35\% of the population are combined compliers, as depicted in Figure \ref{fig:weighting_twins}. This means that at least 35\% of the women have an additional child because of twins or same-sex preferences.

\begin{figure}[!p]
    \centering
    \includegraphics[width=0.8\textwidth]{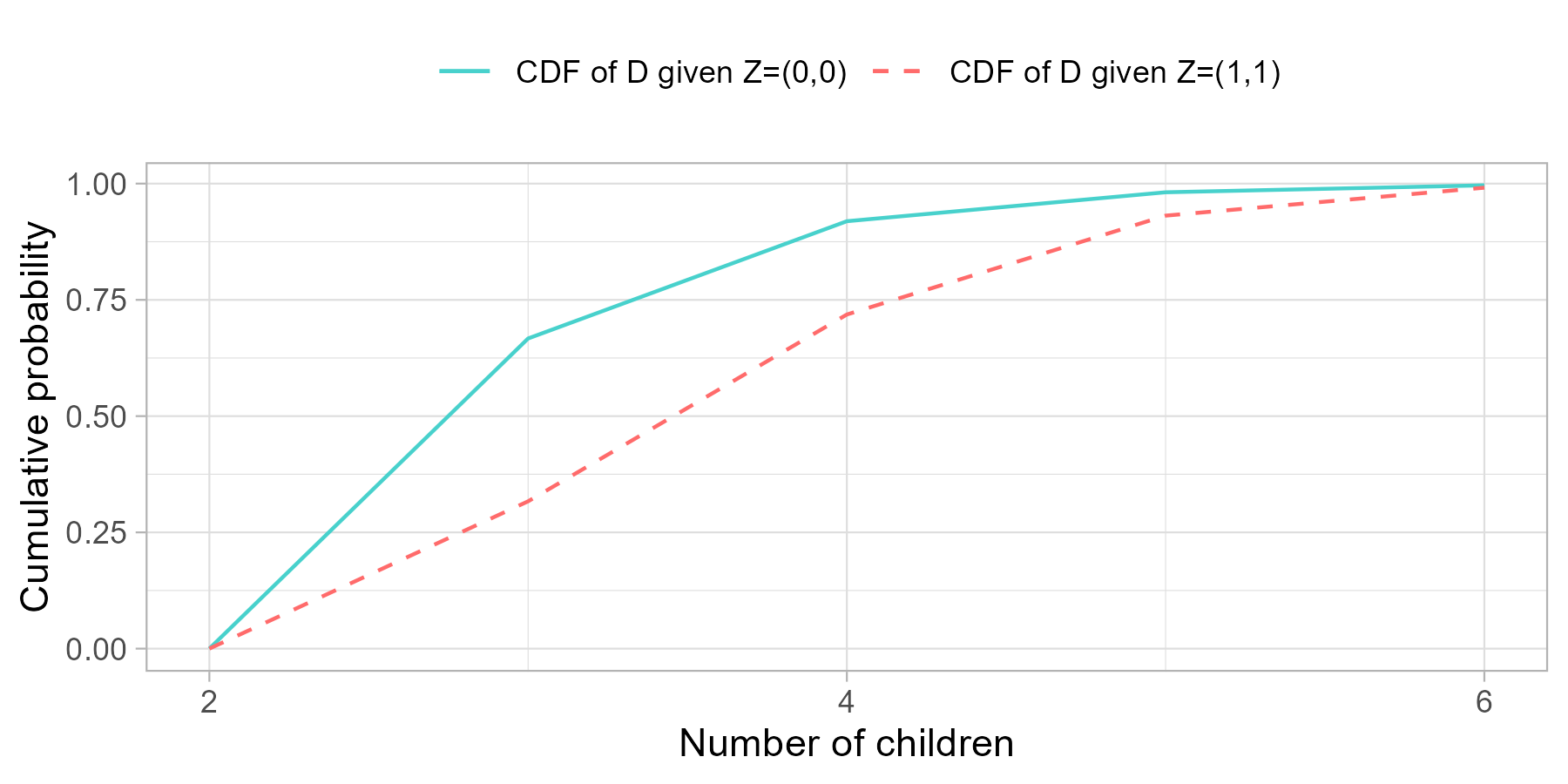}
    \caption{CDFs of the treatment conditional on the outer support of the instrument values in \citeauthor{angrist1996children}'s (\citeyear{angrist1996children}) application. 
    The CDFs  do not cross, indicating that the necessary condition for LiM holds at all treatment levels and no violation of the LiM assumption is detected visually in the full sample.
    }
    \label{fig:cdf_lim_twins}
\end{figure}

\bigskip

\subsubsection{Plausibility of LiM}

Consider the two example types in Table \ref{tab:lim_vs_pm_child}. A woman of type \( t_{3,2,3,3} \) has a strong preference for two girls: she has three children if the first two are a boy and a girl (\( D^{00}=3 \)), two children if the first two are girls (\( D^{01}=2 \)), and three children if twins are born at the second birth (\( D^{10}=3 \) and \( D^{11}=3 \)). Another woman, \( t_{2,3,3,3} \), prefers the first two children to be of different sexes: she has two children if the first two are of mixed sex (\( D^{00}=2 \)), and three children if the first two are of the same sex (\( D^{01}=3 \)). PM does not hold here because it requires either \( P(D^{10} \geq D^{00})=1 \) or \( P(D^{10} \leq D^{00})=1 \).

\begin{table}[tbp]
    \centering
    \caption{Two example types in the \citet{angrist1996children} application. 
    PM rules out one of these two types. 
    Both types are consistent with LiM ($P(D^{11}\geq D^{00})=1$). $Z_1$ is the twinning instrument, and $Z_2$ the same-sex instrument.
    }
    \label{tab:lim_vs_pm_child}
    \begin{tabular}{lcccc} \toprule
        Type & $D^{00}$ & $D^{01}$ & $D^{10}$ & $D^{11}$ \\ \hline
        $t_{3,2,3,3}$ & 3 & 2 & 3 & 3 \tabularnewline
        $t_{2,3,3,3}$ & 2 & 3 & 3 & 3 \\ \bottomrule
    \end{tabular}
\end{table}

\bigskip

\subsubsection{LiM test}

In the study by \citet{angrist1996children}, it is very unlikely that LiM is violated, as one cannot defy the twinning instrument, which always pushes towards compliance (i.e., having an additional child).\footnote{Note that \citet{lewbel2022limited} provide a related discussion in case of a binary treatment.} Consistent with this, Figure \ref{fig:cdf_lim_twins} visually provides no evidence of a LiM violation. Unsurprisingly, the local LiM test (see Section \ref{sec:testing_lim_local}) does not detect a violation.\footnote{Configuration settings for the LiM test are the same as those specified in footnote \ref{note1}.}

\bigskip

\section{Conclusion}
\label{sec:discussion}

This study advances the understanding of causal inference in the context of discrete, ordered and continuous treatments with multiple instruments by introducing a novel approach to identification and estimation. The central theoretical contribution, the CC-ACR, provides an intuitive and robust alternative to the conventional TSLS methodology. Unlike TSLS, the CC-ACR accommodates a less restrictive LiM assumption and offers a clearer interpretation of causal effects by leveraging a weighting scheme directly tied to the shares of combined complier types. This approach is particularly valuable for estimating the average causal effect of a one-level increase in treatment. 

The study also introduces a practical test for the LiM assumption, using causal forests to detect local violations in a data-driven manner. This test enhances the robustness of empirical findings by identifying potential subgroups where the LiM assumption may not hold.

Empirical applications underscore the practical advantages of the CC-ACR framework under the LiM assumption. In estimating the returns to education, the CC-ACR allows for more granular insights into the impact of each additional year of schooling on wages, surpassing the binary approach that only considers college attendance. By leveraging both strong and weak instruments, this method delivers more precise estimates and policy-relevant insights. Similarly, in the analysis of the effect of additional children on female labor market outcomes, the CC-ACR indicates heterogeneity in labor market preferences, facilitated by incorporating the novel twinning instrument. The CC-ACR estimates are obtained by adapting the DML methodology of \citet{chernozhukov2018double} to the setting with discrete, ordered and continuous treatments and multiple instruments.

The present study also reveals several promising avenues for future research. For instance, it is straightforward to extend the results to fuzzy regression discontinuity designs with a multivalued treatment and multiple running variables. Additionally, future research could assess potential power improvements for the LiM test. Techniques such as pruning or replacing the Bonferroni correction with a multiplier bootstrap approach, in the spirit of \citet{huber2022testing}, hold potential.
Moreover, while the CC-ACR represents a weighted average of causal effects, one might instead be interested in obtaining the causal effect of a one-level increase, $(Y^j - Y^{j-1})$, for some specific treatment level $j \in \{ 1,\dots,J\}$. Building upon the work of \citet{kitagawa2021identification} and \citet{huber2017sharp}, partial identification could be explored for the average causal response resulting from, for example, a one-level increase in the treatment level for various combined complier groups. 
Note that for point-identifying the effect along specific treatment margins for a combined complier population, an instrument that pushes individuals towards compliance at that specific margin is required (see, for instance, \citet{bhuller20222sls}).

\bigskip

\bigskip
\bibliography{references}

\newpage
\begin{appendices}

\section{Complete table with response types}
\label{app:full_table_three_valued}

\begin{longtable}{llccccccc}

\caption{Complete table with response types in case of three-valued treatment, $D \in \{0,1,2 \}$, and two instruments, $Z_1$ and $Z_2$. Suppose that the instrument support $\mathcal{Z}= \{ z_0, z_1, z_2, z_3\}$ is ordered such that $E(D|Z=z_0)<E(D|Z=z_1)<E(D|Z=z_2)<E(D|Z=z_3)$ and label the ordered elements as $z_0$, $z_1$, $z_2$, $z_3$. Here, suppose $\mathcal{Z}= \{ z_0, z_1, z_2, z_3\} =\{ (0,0), (0,1), (1,0), (1,1) \}$. \checkmark indicates the response types under the different forms of the monotonicity assumption.} 
\label{tab:all_types_3_values} \\

\hline \multicolumn{1}{c}{} & \multicolumn{1}{c}{} & \multicolumn{1}{c}{$D^{z_0}$} & \multicolumn{1}{c}{$D^{z_1}$} & \multicolumn{1}{c}{$D^{z_2}$}  & \multicolumn{1}{c}{$D^{z_3}$}  & \multicolumn{1}{c}{} & \multicolumn{1}{c}{} & \multicolumn{1}{c}{}  \\
\multicolumn{1}{c}{Combined type} & \multicolumn{1}{c}{Type} & \multicolumn{1}{c}{$D^{00}$} & \multicolumn{1}{c}{$D^{01}$} & \multicolumn{1}{c}{$D^{10}$}  & \multicolumn{1}{c}{$D^{11}$}  & \multicolumn{1}{c}{LiM} & \multicolumn{1}{c}{PM} & \multicolumn{1}{c}{IAM}  \\\hline 
\endfirsthead

\multicolumn{8}{c}
{{ \tablename\ \thetable{} -- continued from previous page}} \\
\hline \multicolumn{1}{c}{Combined type} & \multicolumn{1}{c}{Type} & \multicolumn{1}{c}{$D^{00}$} & \multicolumn{1}{c}{$D^{10}$} & \multicolumn{1}{c}{$D^{01}$}  & \multicolumn{1}{c}{$D^{11}$}  & \multicolumn{1}{c}{LiM} & \multicolumn{1}{c}{PM} & \multicolumn{1}{c}{IAM}  \\ \hline 
\endhead

\hline \multicolumn{8}{r}{{Continued on next page}} \\ 
\endfoot

\hline 
\endlastfoot

    $cn_{2,2}$ & $n_{2,2,2,2}$ & 2 & 2 & 2 & 2 & \checkmark & \checkmark & \checkmark \\
     & $n_{2,1,2,2}$ &  2 & 1 & 2 & 2 & \checkmark &  &  \\
     & $n_{2,0,2,2}$ &  2 & 0 & 2 & 2 & \checkmark &  &  \\
     & $n_{2,2,1,2}$ &  2 & 2 & 1 & 2 & \checkmark &  &  \\
     & $n_{2,1,1,2}$ &  2 & 1 & 1 & 2 & \checkmark &  &  \\
     & $n_{2,0,1,2}$ &  2 & 0 & 1 & 2 & \checkmark &  &  \\
     & $n_{2,2,0,2}$ &  2 & 2 & 0 & 2 & \checkmark &  &  \\
     & $n_{2,1,0,2}$ &  2 & 1 & 0 & 2 & \checkmark &  &  \\
     & $n_{2,0,0,2}$ &  2 & 0 & 0 & 2 & \checkmark &  &  \\ \hline
    $cc_{1,2}$ & $c_{1,2,2,2}$ & 1 & 2 & 2 & 2 & \checkmark & \checkmark & \checkmark \\
     & $c_{1,1,2,2}$ &  1 & 1 & 2 & 2 & \checkmark & \checkmark & \checkmark \\
     & $c_{1,0,2,2}$ &  1 & 0 & 2 & 2 & \checkmark &  &  \\
     & $c_{1,2,1,2}$ &  1 & 2 & 1 & 2 & \checkmark & \checkmark &  \\
     & $c_{1,1,1,2}$ &  1 & 1 & 1 & 2 & \checkmark & \checkmark & \checkmark \\
     & $c_{1,0,1,2}$ &  1 & 0 & 1 & 2 & \checkmark &  &  \\
     & $c_{1,2,0,2}$ &  1 & 2 & 0 & 2 & \checkmark &  &  \\
     & $c_{1,1,0,2}$ &  1 & 1 & 0 & 2 & \checkmark &  &  \\
     & $c_{1,0,0,2}$ &  1 & 0 & 0 & 2 & \checkmark &  &  \\ \hline
    $cc_{0,2}$ & $c_{0,2,2,2}$ & 0 & 2 & 2 & 2 & \checkmark & \checkmark & \checkmark \\
     & $c_{0,1,2,2}$ &  0 & 1 & 2 & 2 & \checkmark & \checkmark & \checkmark \\
     & $c_{0,0,2,2}$ &  0 & 0 & 2 & 2 & \checkmark & \checkmark & \checkmark \\
     & $c_{0,2,1,2}$ &  0 & 2 & 1 & 2 & \checkmark & \checkmark &  \\
     & $c_{0,1,1,2}$ &  0 & 1 & 1 & 2 & \checkmark & \checkmark & \checkmark \\
     & $c_{0,0,1,2}$ &  0 & 0 & 1 & 2 & \checkmark & \checkmark & \checkmark \\
     & $c_{0,2,0,2}$ &  0 & 2 & 0 & 2 & \checkmark & \checkmark &  \\
     & $c_{0,1,0,2}$ &  0 & 1 & 0 & 2 & \checkmark & \checkmark &  \\
     & $c_{0,0,0,2}$ &  0 & 0 & 0 & 2 & \checkmark & \checkmark & \checkmark \\ \hline
    $cd_{2,1}$ & $d_{2,2,2,1}$ & 2 & 2 & 2 & 1 &  &  &  \\
     & $d_{2,1,2,1}$ &  2 & 1 & 2 & 1 &  &  &  \\
     & $d_{2,0,2,1}$ &  2 & 0 & 2 & 1 &  &  &  \\
     & $d_{2,2,1,1}$ &  2 & 2 & 1 & 1 &  &  &  \\
     & $d_{2,1,1,1}$ &  2 & 1 & 1 & 1 &  &  &  \\
     & $d_{2,0,1,1}$ &  2 & 0 & 1 & 1 &  &  &  \\
     & $d_{2,2,0,1}$ &  2 & 2 & 0 & 1 &  &  &  \\
     & $d_{2,1,0,1}$ &  2 & 1 & 0 & 1 &  &  &  \\
     & $d_{2,0,0,1}$ &  2 & 0 & 0 & 1 &  &  &  \\ \hline
    $cn_{1,1}$ & $n_{1,2,2,1}$ & 1 & 2 & 2 & 1 & \checkmark &  &  \\
     & $n_{1,1,2,1}$ &  1 & 1 & 2 & 1 & \checkmark &  &  \\
     & $n_{1,0,2,1}$ &  1 & 0 & 2 & 1 & \checkmark &  &  \\
     & $n_{1,2,1,1}$ &  1 & 2 & 1 & 1 & \checkmark &  &  \\
     & $n_{1,1,1,1}$ &  1 & 1 & 1 & 1 & \checkmark & \checkmark & \checkmark \\
     & $n_{1,0,1,1}$ &  1 & 0 & 1 & 1 & \checkmark &  &  \\
     & $n_{1,2,0,1}$ &  1 & 2 & 0 & 1 & \checkmark &  &  \\
     & $n_{1,1,0,1}$ &  1 & 1 & 0 & 1 & \checkmark &  &  \\
     & $n_{1,0,0,1}$ &  1 & 0 & 0 & 1 & \checkmark &  &  \\ \hline
    $cc_{0,1}$ & $c_{0,2,2,1}$ & 0 & 2 & 2 & 1 & \checkmark &  &  \\
     & $c_{0,1,2,1}$ &  0 & 1 & 2 & 1 & \checkmark &  &  \\
     & $c_{0,0,2,1}$ &  0 & 0 & 2 & 1 & \checkmark &  &  \\
     & $c_{0,2,1,1}$ &  0 & 2 & 1 & 1 & \checkmark &  &  \\
     & $c_{0,1,1,1}$ &  0 & 1 & 1 & 1 & \checkmark & \checkmark & \checkmark \\
     & $c_{0,0,1,1}$ &  0 & 0 & 1 & 1 & \checkmark & \checkmark & \checkmark \\
     & $c_{0,2,0,1}$ &  0 & 2 & 0 & 1 & \checkmark &  &  \\
     & $c_{0,1,0,1}$ &  0 & 1 & 0 & 1 & \checkmark & \checkmark &  \\
     & $c_{0,0,0,1}$ &  0 & 0 & 0 & 1 & \checkmark & \checkmark & \checkmark \\ \hline
    $cd_{2,0}$ & $d_{2,2,2,0}$ & 2 & 2 & 2 & 0 &  &  &  \\
     & $d_{2,1,2,0}$ &  2 & 1 & 2 & 0 &  &  &  \\
     & $d_{2,0,2,0}$ &  2 & 0 & 2 & 0 &  &  &  \\
     & $d_{2,2,1,0}$ &  2 & 2 & 1 & 0 &  &  &  \\
     & $d_{2,1,1,0}$ &  2 & 1 & 1 & 0 &  &  &  \\
     & $d_{2,0,1,0}$ &  2 & 0 & 1 & 0 &  &  &  \\
     & $d_{2,2,0,0}$ &  2 & 2 & 0 & 0 &  &  &  \\
     & $d_{2,1,0,0}$ &  2 & 1 & 0 & 0 &  &  &  \\
     & $d_{2,0,0,0}$ &  2 & 0 & 0 & 0 &  &  &  \\ \hline
    $cd_{1,0}$ & $d_{1,2,2,0}$ & 1 & 2 & 2 & 0 &  &  &  \\
     & $d_{1,1,2,0}$ &  1 & 1 & 2 & 0 &  &  &  \\
     & $d_{1,0,2,0}$ &  1 & 0 & 2 & 0 &  &  &  \\
     & $d_{1,2,1,0}$ &  1 & 2 & 1 & 0 &  &  &  \\
     & $d_{1,1,1,0}$ &  1 & 1 & 1 & 0 &  &  &  \\
     & $d_{1,0,1,0}$ &  1 & 0 & 1 & 0 &  &  &  \\
     & $d_{1,2,0,0}$ &  1 & 2 & 0 & 0 &  &  &  \\
     & $d_{1,1,0,0}$ &  1 & 1 & 0 & 0 &  &  &  \\
     & $d_{1,0,0,0}$ &  1 & 0 & 0 & 0 &  &  &  \\ \hline
    $cn_{0,0}$ & $n_{0,2,2,0}$ & 0 & 2 & 2 & 0 & \checkmark &  &  \\
     & $n_{0,1,2,0}$ &  0 & 1 & 2 & 0 & \checkmark &  &  \\
     & $n_{0,0,2,0}$ &  0 & 0 & 2 & 0 & \checkmark &  &  \\
     & $n_{0,2,1,0}$ &  0 & 2 & 1 & 0 & \checkmark &  &  \\
     & $n_{0,1,1,0}$ &  0 & 1 & 1 & 0 & \checkmark &  &  \\
     & $n_{0,0,1,0}$ &  0 & 0 & 1 & 0 & \checkmark &  &  \\
     & $n_{0,2,0,0}$ &  0 & 2 & 0 & 0 & \checkmark &  &  \\
     & $n_{0,1,0,0}$ &  0 & 1 & 0 & 0 & \checkmark &  &  \\
     & $n_{0,0,0,0}$ &  0 & 0 & 0 & 0 & \checkmark & \checkmark & \checkmark \\ 
\end{longtable}

\bigskip

\newpage
\section{Proofs}
\label{app:proofs}

\subsection{Proof of Theorem 1}
\label{app:proof_theorem1}

First note that $\sum_{k, l} P(T=cc_{k,l}) = \sum_{k \leq l} P(T=cc_{k,l}) = 1$ because of LiM. 
Then, consider the first part of the numerator of $\beta \equiv \frac{E(Y|Z_1=Z_2=...=Z_K=1) - E(Y|Z_1=Z_2=...=Z_K=0)}{E(D|Z_1=Z_2=...=Z_K=1) - E(D|Z_1=Z_2=...=Z_K=0)}$:

\begin{equation*}
    \begin{aligned}
        E(Y|Z_1=Z_2=...=Z_K=1) &= E(Y|\widetilde{Z}=1) \\  
        &= \sum_{k, l} E(Y|\widetilde{Z}=1,T=cc_{k,l}) P(T=cc_{k,l}|\widetilde{Z}=1) \\
        &= \sum_{k \leq l} E(Y|\widetilde{Z}=1,T=cc_{k,l}) P(T=cc_{k,l}|\widetilde{Z}=1) \\
        &= \sum_{k \leq l} E(Y^l|T=cc_{k,l}) P(T = cc_{k,l}),
    \end{aligned}
\end{equation*}
where the third equality follows from LiM and the last equality follows from the exclusion and unconfoundedness/independence assumptions.

Similarly, consider the other components of 
\begin{equation*}
    E(Y|Z_1=Z_2=...=Z_K=0) = E(Y|\widetilde{Z}=0) = \sum_{k \leq l} E(Y^k|T=c_{k,l}) P(T=cc_{k,l}),
\end{equation*}
and 
\begin{equation*}
    E(D|Z_1=Z_2=...=Z_K=1) = E(D|\widetilde{Z}=1) = \sum_{k \leq l} l \cdot P(T = cc_{k,l}), 
\end{equation*}
and
\begin{equation*}
    E(D|Z_1=Z_2=...=Z_K=0) = E(D|\widetilde{Z}=0) = \sum_{k \leq l} k \cdot P(T = cc_{k,l}). 
\end{equation*}
\bigskip

\noindent Combining the above results:
\begingroup
\allowdisplaybreaks
    \begin{align*}
        & \frac{E(Y|\widetilde{Z}=1) - E(Y|\widetilde{Z}=0)}{E(D|\widetilde{Z}=1) - E(D|\widetilde{Z}=0)} \\ 
        &= \frac{\sum_{k \leq l} E(Y^l|T=cc_{k,l}) P(T= cc_{k,l}) - \sum_{k \leq l} E(Y^k|T = cc_{k,l}) P(T = cc_{k,l})}{\sum_{k \leq l} l \cdot P(T = cc_{k,l}) - \sum_{k \leq l} k \cdot P(T = cc_{k,l})} \\
        &= \frac{\sum_{k \leq l} E(Y^l - Y^k|T=cc_{k,l}) P(T= cc_{k,l})}{\sum_{k \leq l} (l-k) \cdot P(T = cc_{k,l})} \\&= \frac{\sum_{k \leq l} E(Y^l - Y^k|T=cc_{k,l}) P(T= cc_{k,l})}{\sum_{k \leq l} (l-k) \cdot P(T = cc_{k,l})} \\
        &= \frac{\sum_{k \leq l} E(Y^l - Y^k|T=cc_{k,l}) P(T= cc_{k,l})}{\sum_{k \leq l} (l-k) \cdot P(T = cc_{k,l})} \\
        &= \frac{\sum_{k < l} E(Y^l - Y^k|T=cc_{k,l}) P(T= cc_{k,l})}{\sum_{k < l} (l-k) \cdot P(T = cc_{k,l})} + \frac{\sum_{k = l} E(Y^l - Y^k|T=cc_{k,l}) P(T= cc_{k,l})}{\sum_{k = l} (l-k) \cdot P(T = cc_{k,l})} \\
        &= \frac{\sum_{k < l} E(Y^l - Y^k|T=cc_{k,l}) P(T= cc_{k,l})}{\sum_{k < l} (l-k) \cdot P(T = cc_{k,l})} \\
        &= \sum_{k < l} \frac{P(T= cc_{k,l})}{\sum_{k < l} (l-k) \cdot P(T = cc_{k,l})} E(Y^l - Y^k|T=cc_{k,l}).
    \end{align*}
\endgroup

\bigskip

\subsection{Proof of Proposition 1}
\label{app:proof_tsls}

Suppose that the treatment $D$ is discrete with bounded support. Denote with M the number of elements in the rectangular instrument support $\mathcal{Z}$ ordered such that $l<m$ implies $E(D|Z=l)<E(D|Z=m)$. Label the ordered elements as $z_1$, $z_2$, ..., $z_M$. 
Theorem 2 of \citet{angrist1995two} establishes that TSLS combined with Assumptions 1 to 3 estimates 
\begin{equation*}
        \beta_{TSLS}  \equiv  \sum_{m=1}^M \mu_m \cdot \beta_{m,m-1}, 
\end{equation*}
where
\begin{equation*}
        \mu_m = (E(D|Z=z_m) - E(D|Z=z_{m-1})) \cdot \frac{\sum_{l=m}^M P(Z=z_l) (E(D|Z=z_l)-E(D))}{\sum_{l=0}^M P(Z=z_l) E(D|Z=z_l)(E(D|Z=z_l)-E(D))} 
\end{equation*}
and
\begin{equation*}
    \beta_{m,m-1} = \frac{E(Y|Z=z_m)-E(Y|Z=z_{m-1})}{E(D|Z=z_m)-E(D|Z=z_{m-1})}.
\end{equation*}
\bigskip

Using this as a starting point, I now show that this can be rewritten to obtain an interpretation of a weighted average of causal responses for different response types:
\begin{equation*}
    \begin{aligned}
        & \beta_{TSLS} \\ 
        &= \sum_{m=1}^M (E(D|Z=z_m) - E(D|Z=z_{m-1})) \cdot \frac{\sum_{l=m}^M P(Z=z_l) (E(D|Z=z_l)-E(D))}{\sum_{l=0}^M P(Z=z_l) E(D|Z=z_l)(E(D|Z=z_l)-E(D))} \\ 
        & \hphantom{=} \cdot \frac{E(Y|Z=z_m)-E(Y|Z=z_{m-1})}{E(D|Z=z_m)-E(D|Z=z_{m-1})} \\
        &= \sum_{m=1}^M \frac{\sum_{l=m}^M P(Z=z_l) (E(D|Z=z_l)-E(D))}{\sum_{l=0}^M P(Z=z_l) E(D|Z=z_l)(E(D|Z=z_l)-E(D))} \cdot E(Y|Z=z_m)-E(Y|Z=z_{m-1}) \\
        &= \sum_{m=1}^M \frac{\sum_{l=m}^M P(Z=z_l) (E(D|Z=z_l)-E(D))}{\sum_{l=0}^M P(Z=z_l) E(D|Z=z_l)(E(D|Z=z_l)-E(D))} \cdot  E(Y^{D^{z_m}}-Y^{D^{z_{m-1}}}) \\
        &\equiv \sum_{m=1}^M \omega_m \cdot E(Y^{D^{z_m}}-Y^{D^{z_{m-1}}}), 
    \end{aligned}
\end{equation*}
where the weights are
\begin{equation*}
    \begin{aligned}
        \omega_m &= \frac{\sum_{l=m}^M P(Z=z_l) (E(D|Z=z_l)-E(D))}{\sum_{l=0}^M P(Z=z_l) E(D|Z=z_l)(E(D|Z=z_l)-E(D))}.
    \end{aligned}
\end{equation*} \bigskip

Denote $\mathcal{T}$ the set of all response types $t$, and $\sum_{t \in \mathcal{T}} P(T=t)=1$. Further denote $I(\cdot)$ the indicator function, which equals one if its argument is true and zero otherwise.
Then, it can be shown that TSLS preserves the interpretation of a weighted average of causal responses $Y^a-Y^b$ where $a>b$:
\allowdisplaybreaks
\begin{align*}
        & \beta_{TSLS} = \sum_{m=1}^M \omega_m  \cdot E(Y^{D^{z_m}}-Y^{D^{z_{m-1}}}) \\
        &= \sum_{m=1}^M \omega_m \left( \sum_{t \in \mathcal{T}} P(T=t) \cdot E(Y^{D^{z_m}}-Y^{D^{z_{m-1}}}|T=t) \right) \\
        &= \sum_{t \in \mathcal{T}}  \left( P(T=t)  \sum_{m=1}^M \omega_m \cdot  E(Y^{D^{z_m}}-Y^{D^{z_{m-1}}}|T=t) \right) \\
        &= \sum_{t \in \mathcal{T}} \Bigl( P(T=t)  \sum_{m=1}^M \Bigl\{ I(D^{z_m} > D^{z_{m-1}}) \cdot \omega_m \cdot  E(Y^{D^{z_m}}-Y^{D^{z_{m-1}}}|T=t)  \\ 
        &  \hphantom{== \sum_{g \in \mathcal{T}} P(T=t)  \sum_{m=1}^M}  - I(D^{z_m} < D^{z_{m-1}}) \cdot \omega_m \cdot  E(Y^{D^{z_{m-1}}}-Y^{D^{z_m}}|T=t) \Bigr\} \Bigr) \\
        & \equiv  \sum_{t \in \mathcal{T}} P(T=t) \sum_{m=1}^M \iota_{m,m-1} \cdot \omega_m  \cdot E(Y^{D^{z_m}}-Y^{D^{z_{m-1}}}|T=t),
\end{align*}
where
\begin{equation*}
  \iota_{m,m-1} \equiv I(D^{z_m}\geq D^{z_{m-1}}) - I(D^{z_m}\leq D^{z_{m-1}}) =
    \begin{cases}
      -1 & \text{ if $D^{z_m} < D^{z_{m-1}}$}\\
      \hphantom{-} 1 & \text{ if $D^{z_m} > D^{z_{m-1}}$}\\
      \hphantom{-} 0 & \text{ if $D^{z_m} = D^{z_{m-1}}$}
    \end{cases},       
\end{equation*}
and
\begin{equation*}
        \omega_m = \frac{\sum_{l=m}^M P(Z=z_l) (E(D|Z=z_l)-E(D))}{\sum_{l=0}^M P(Z=z_l) E(D|Z=z_l)(E(D|Z=z_l)-E(D))}.
\end{equation*}
\bigskip

\noindent The fourth equality holds since $Y^{D^{z_m}}-Y^{D^{z_{m-1}}}=0$ when $D^{z_m} = D^{z_{m-1}}$.
Note that the numerator of $\omega_m$ can be re-written as follows:
\begin{align*}
    & \sum_{l=m}^M P(Z=z_l) (E(D|Z=z_l)-E(D)) \\ 
    &= \sum_{l=m}^M \left( P(Z=z_l) E(D|Z=z_l) \right) - \sum_{l=m}^M \left( P(Z=z_l) E(D) \right) \\
    &= \sum_{l=m}^M \left( P(Z=z_l) E(D|Z=z_l) \right) - E(D)  \sum_{l=m}^M P(Z=z_l) \\
    &= \sum_{l=m}^M \left( P(Z=z_l) E(D|Z=z_l) \right) - P(Z\geq k ) E(D) \\
    &= \sum_{l=m}^M \left( P(Z=z_l) E(D|Z=z_l) \right) - P(Z\geq z_k ) (P(Z<z_k) E(D|Z<z_k) + P(Z\geq z_k) E(D|Z \geq z_k)) \\
    &= (1-P(Z\geq z_k)) P(Z \geq z_k) E(D|Z\geq z_k) - P(Z\geq z_k) P(Z<z_k) E(D|Z<z_k)\\
    &= (1-P(Z\geq z_k)) P(Z \geq z_k) E(D|Z\geq z_k) - P(Z \geq z_k) (1-P(Z\geq z_k))  E(D|Z<z_k) \\
    &= (1-P(Z\geq z_k)) P(Z \geq z_k) \cdot \left\{ E(D|Z\geq z_k) - E(D|Z<z_k) \right\}.
\end{align*}

\bigskip

\subsection{Proof of alternative formulation of Theorem 1}
\label{app:proof_theorem1_alternative}

\noindent Write Y as follows:
\begin{equation*}
    \begin{aligned}
        Y &= I(Z_1=Z_2=...=Z_K=1) \cdot Y^{D^{1...1...1}} + I(Z_1=Z_2=...=Z_K=0) \cdot Y^{D^{0...0...0}} \\
        & \ \ \ \ + I(Z_1=q, ..., Z_k=r,...,Z_K=s) \cdot Y^{D^{q...r...s}} \\
        &= \left(I(Z_1=Z_2=...=Z_K=1) \cdot \sum_{j=0}^J Y^j \cdot I(D^{1...1...1} \geq j) \right) \\
        & \ \ \ \ + \left(I(Z_1=Z_2=...=Z_K=0) \cdot \sum_{j=0}^J Y^j \cdot I(D^{0...0...0} \geq j) \right)
        \\
        & \ \ \ \ + \left(I(Z_1=q, ..., Z_k=r,...,Z_K=s) \cdot \sum_{j=0}^J Y^j \cdot I(D^{q...r...s} \geq j) \right), 
    \end{aligned}
\end{equation*}
\bigskip
$\forall q,r,s$ such that $q \neq r \neq s$.

\noindent First, consider the numerator in $\beta_{\text{CC-ACR}} \equiv \frac{E(Y|Z_1=Z_2=...=Z_K=1) - E(Y|Z_1=Z_2=...=Z_K=0)}{E(D|Z_1=Z_2=...=Z_K=1) - E(D|Z_1=Z_2=...=Z_K=0)}$:
\begin{equation*}
    \begin{aligned}
       & E(Y|Z_1=Z_2=...=Z_K=1) - E(Y|Z_1=Z_2=...=Z_K=0) \\ 
    &= E\left(\sum_{j=0}^J Y^j \cdot I(D^{1...1...1} \geq j)|Z_1=Z_2=...=Z_K=1\right) \\ & 
    \ \ \ \ -E\left(\sum_{j=0}^J Y^j \cdot I(D^{0...0...0} \geq j)|Z_1=Z_2=...=Z_K=0\right) \\
        & = E \Biggl( \sum_{j=0}^J Y^j \cdot (I(D^{1...1...1} \geq j)-I(D^{0...0...0} \geq j)) \Biggr) \\
    &= E \Biggl( \sum_{j=0}^J Y^j \cdot (I(D^{1...1...1} \geq j)-I(D^{1...1...1} \geq j+1) -I(D^{0...0...0} \geq j) - I(D^{0...0...0} \geq j+1)) \Biggr) \\
    &= E \Biggl( Y_0 \cdot (I(D^{1...1...1} \geq 0)-I(D^{1...1...1} \geq 1) -I(D^{0...0...0} \geq 0) - I(D^{0...0...0} \geq 1))  \\
    &\hphantom{=} + \sum_{j=1}^J Y^j \cdot (I(D^{1...1...1} \geq j)-I(D^{1...1...1} \geq j+1) -I(D^{0...0...0} \geq j) - I(D^{0...0...0} \geq j+1)) \Biggr) \\
    &= E\Biggl(Y_0 \cdot (I(D^{1...1...1} \geq 0)-I(D^{0...0...0} \geq 0)) + \sum_{j=1}^J (Y^j - Y^{j-1}) \cdot (I(D^{1...1...1} \geq j)-I(D^{0...0...0} \geq j)) \Biggr) \\
    &= E\Biggl(\sum_{j=1}^J (Y^j - Y^{j-1}) \cdot (I(D^{1...1...1} \geq j)-I(D^{0...0...0} \geq j)) \Biggr).
    \end{aligned}
\end{equation*}

\noindent $I(D^{1...1...1} \geq j )-I(D^{0...0...0} \geq j )$ equals zero or one since $I(D^{1...1...1} \geq j ) \geq I(D^{0...0...0} \geq j )$. \\

\noindent Subsequently:
\begin{equation*}
    \begin{aligned}
        & \sum_{j=1}^J E(Y^j - Y^{j-1} | I(D^{1...1...1} \geq j )-I(D^{0...0...0} \geq j )=1) \cdot P(I(D^{1...1...1} \geq j )-I(D^{0...0...0} \geq j )=1) \\
        &= \sum_{j=1}^J E(Y^j - Y^{j-1} | D^{1...1...1} \geq j > D^{0...0...0} ) \cdot P(D^{1...1...1} \geq j > D^{0...0...0} ).
    \end{aligned}
\end{equation*} 

Now, write $D$ as follows:
\begin{equation*}
    \begin{aligned}
        D &= I(Z_1=Z_2=...=Z_K=1) \cdot D^{1...1...1} + I(Z_1=Z_2=...=Z_K=0) \cdot D^{0...0...0} \\
        &  \ \ \ \ + I(Z_1=q, ..., Z_k=r,...,Z_K=s) \cdot D^{q...r...s} \\
        &= \left(I(Z_1=Z_2=...=Z_K=1) \cdot \sum_{j=0}^J j \cdot I(D^{1...1...1} \geq j) \right) \\
        & \ \ \ \ + 
        \left(I(Z_1=Z_2=...=Z_K=0) \cdot \sum_{j=0}^J j \cdot I(D^{0...0...0} \geq j) \right)\\
        &  \ \ \ \ + 
        \left(I(Z_1=q, ..., Z_k=r,...,Z_K=s) \cdot \sum_{j=0}^J j \cdot I(D^{q...r...s} \geq j) \right),
    \end{aligned}
\end{equation*}
$\forall q,r,s$ such that $q \neq r \neq s$.

Then, consider the denominator in $\beta_{\text{CC-ACR}} \equiv \frac{E(Y|Z_1=Z_2=...=Z_K=1) - E(Y|Z_1=Z_2=...=Z_K=0)}{E(D|Z_1=Z_2=...=Z_K=1) - E(D|Z_1=Z_2=...=Z_K=0)}$:
\allowdisplaybreaks
\begin{align*}
        & E(D|Z_1=Z_2=...=Z_K=1) - E(D|Z_1=Z_2=...=Z_K=0) \\ &= E\left(\sum_{j=0}^J j \cdot I(D^{1...1...1} \geq j)|Z_1=Z_2=...=Z_K=1\right) \\
        & \ \ \ \ -E\left(\sum_{j=0}^J j \cdot I(D^{0...0...0} \geq j)|Z_1=Z_2=...=Z_K=0\right) \\
        &= E\left(\sum_{j=0}^J j \cdot (I(D^{1...1...1}=j)-I(D^{0...0...0}=j)) \right) \\
        &= E\left(\sum_{j=0}^J j \cdot (I(D^{1...1...1} \geq j)-I(D^{1...1...1} \geq j+1) -I(D^{0...0...0} \geq j) - I(D^{0...0...0} \geq j+1))\right) \\
        &= E\left(\sum_{j=1}^J I(D^{1...1...1} \geq j)-I(D^{0...0...0} \geq j) \right) \\
        &= \sum_{j=1}^J P(D^{1...1...1} \geq j > D^{0...0...0}).
\end{align*}

\noindent It is required that $P(D^{1...1...1} \geq j > D^{0...0...0})>0$ for some $j$ which imposes a relevance assumption on the instrument. Moreover, $P(D^{1...1...1} \geq l > D^{0...0...0})= \sum_{l>k} P(T=c_{l,k})$.

\noindent Then:
\begin{equation*}
    \begin{aligned}
        & \frac{E(Y|Z_1=Z_2=...=Z_K=1) - E(Y|Z_1=Z_2=...=Z_K=0)}{E(D|Z_1=Z_2=...=Z_K=1) - E(D|Z_1=Z_2=...=Z_K=0)} \\
        &= \sum_{j=1}^J \frac{P(D^{1...1...1} \geq j > D^{0...0...0})}{\sum_{i=1}^J P(D^{1...1...1} \geq i > D^{0...0...0})} E(Y^j - Y^{j-1} | D^{1...1...1} \geq j > D^{0...0...0}) \\
        &=  \sum_{j=1}^J \frac{P(T=c_{l,k})}{\sum_{i=1}^J  \sum_{l>i} P(T=c_{l,i})} E(Y^j - Y^{j-1} | D^{1...1...1} \geq j > D^{0...0...0}).
    \end{aligned}
\end{equation*}

\bigskip

\subsection{Proof of alternative representation of the TSLS estimand}
\label{app:proof_tsls_alternative}

Write Theorem 2 of \citet{angrist1995two} as follows:
\begin{equation}\label{eq:theorem2_ia1994}
        \beta_{TSLS}  \equiv  \sum_{m=1}^M \delta_{m,m-1} \cdot \omega_m \cdot \beta_{m,m-1}, 
\end{equation}
where
\begin{equation*}
    \delta_{m,m-1} = E(D|Z=z_m) - E(D|Z=z_{m-1}),
\end{equation*}
and
\begin{equation*}
        \omega_m = \frac{\sum_{l=m}^M P(Z=z_l) (E(D|Z=z_l)-E(D))}{\sum_{l=0}^M P(Z=z_l) E(D|Z=z_l)(E(D|Z=z_l)-E(D))},
\end{equation*}
and
\begin{equation*}
    \beta_{m,m-1} = \frac{E(Y|Z=z_m)-E(Y|Z=z_{m-1})}{E(D|Z=z_m)-E(D|Z=z_{m-1})}.
\end{equation*}
\bigskip

\noindent It holds that 
\begin{equation*}
    \begin{aligned}
        E(Y|Z=z_m)-E(Y|Z=z_{m-1}) &= \sum_{j=1}^J  P(D^{z_m}\geq j > D^{z_{m-1}}) \cdot E(Y^j - Y^{j-1}| D^{z_m}\geq j > D^{z_{m-1}})\\
        &\hphantom{=} + \sum_{j=1}^J  P(D^{z_m} < j \leq D^{z_{m-1}}) \cdot E(Y^j - Y^{j-1}| D^{z_m} < j \leq D^{z_{m-1}}).
    \end{aligned}
\end{equation*}

\noindent This can be seen as follows: 
\begingroup
\allowdisplaybreaks
    \begin{align*}
    & E(Y|Z=z_m) - E(Y|Z=z_{m-1}) \\ 
    &= E \left( \sum_{j=0}^J Y^j \cdot I(D^{z_m} \geq j)|Z=z_m \right)-E \left(\sum_{j=0}^J Y^j \cdot I(D^{z_{m-1}} \geq j)|Z=z_{m-1} \right) \\
    &= E \left(\sum_{j=0}^J Y^j \cdot (I(D^{z_m} \geq j)-I(D^{z_{m-1}} \geq j)) \right) \\
    &= E \left(\sum_{j=0}^J Y^j \cdot (I(D^{z_m} \geq j)-I(D^{z_m} \geq j+1) -I(D^{z_{m-1}} \geq j) - I(D^{z_{m-1}} \geq j+1)) \right) \\
    &= E \Biggl( Y^{j-1} \cdot (I(D^{z_m} \geq 0)-I(D^{z_m} \geq 1) -I(D^{z_{m-1}} \geq j-1) - I(D^{z_{m-1}} \geq 1))    \\
    &\hphantom{=} +  \sum_{j=1}^J Y^j \cdot (I(D^{z_m} \geq j)-I(D^{z_m} \geq j+1) -I(D^{z_{m-1}} \geq j) - I(D^{z_{m-1}} \geq j+1)) \Biggr) \\
    &= E \Biggl( Y^{j-1} \cdot (I(D^{z_m} \geq j-1)-I(D^{z_{m-1}} \geq j-1)) \\
    &\hphantom{=}  + \sum_{j=1}^J (Y^j - Y^{j-1}) \cdot (I(D^{z_m} \geq j)-I(D^{z_{m-1}} \geq j)) \Biggr) \\
    &= E(\sum_{j=1}^J (Y^j - Y^{j-1}) \cdot (I(D^{z_m} \geq j)-I(D^{z_{m-1}} \geq j)) )\\
    &= \sum_{j=1}^J E( (Y^j - Y^{j-1}) \cdot (I(D^{z_m} \geq j)-I(D^{z_{m-1}} \geq j)) ) \\
    &= \sum_{j=1}^J  E\left( E( (Y^j - Y^{j-1}) \cdot (I(D^{z_m} \geq j)-I(D^{z_{m-1}} \geq j)) | I(D^{z_m} \geq j)-I(D^{z_{m-1}} \geq j) ) \right) \\
    &= \sum_{j=1}^J \Bigl\{ 1 \cdot P(I(D^{z_m} \geq j)-I(D^{z_{m-1}} \geq j) =1) \\ 
    & \qquad \cdot E( (Y^j - Y^{j-1}) \cdot (I(D^{z_m} \geq j)-I(D^{z_{m-1}} \geq j)) | I(D^{z_m} \geq j)-I(D^{z_{m-1}} \geq j) =1 )\\
    &\hphantom{=} + 0 \cdot P(I(D^{z_m} \geq j)-I(D^{z_{m-1}} \geq j) =0) \\ 
    & \qquad \cdot E( (Y^j - Y^{j-1}) \cdot (I(D^{z_m} \geq j)-I(D^{z_{m-1}} \geq j)) | I(D^{z_m} \geq j)-I(D^{z_{m-1}} \geq j) =0 )\\
    &\hphantom{=} -1 \cdot P(I(D^{z_m} \geq j)-I(D^{z_{m-1}} \geq j) =-1) \\
    & \qquad \cdot E( (Y^j - Y^{j-1}) \cdot (I(D^{z_m} \geq j)-I(D^{z_{m-1}} \geq j)) | I(D^{z_m} \geq j)-I(D^{z_{m-1}} \geq j) =-1 ) \Bigr\}\\
    &= 1 \cdot \sum_{j=1}^J P(D^{z_m}\geq j > D^{z_{m-1}}) \cdot E( (Y^j - Y^{j-1}) \cdot 1 | D^{z_m}\geq j > D^{z_{m-1}})\\
    &\hphantom{=} -1 \cdot \sum_{j=1}^J P(D^{z_m} < j \leq D^{z_{m-1}}) \cdot E((Y^j - Y^{j-1}) \cdot (-1) | D^{z_m} < j \leq D^{z_{m-1}} )\\
    &= \sum_{j=1}^J P(D^{z_m}\geq j > D^{z_{m-1}}) \cdot E(Y^j - Y^{j-1} | D^{z_m}\geq j > D^{z_{m-1}})\\
    &\hphantom{=} - \sum_{j=1}^J P(D^{z_{m-1}} \geq j > D^{z_m}) \cdot E(Y^{j-1} - Y^j | D^{z_{m-1}} \geq j > D^{z_m} ).
    \end{align*}
\endgroup

\noindent Now, it rests to show that 
\begin{equation*}
    E(D|Z=z_m)-E(D|Z=z_{m-1}) = \sum_{i=1}^J P(D^{z_m} \geq i > D^{z_{m-1}}).
\end{equation*}

\noindent This can be shown as follows:
\begin{equation*}
    \begin{aligned}
        D = Z \cdot D^{z_m} + (1-Z) D^{z_{m-1}} 
        &= \left(Z \cdot \sum_{j=0}^J j \cdot I(D^{z_m} \geq j) \right) + 
        \left((1-Z) \cdot \sum_{j=0}^J j \cdot I(D^{z_{m-1}} \geq j)\right).
    \end{aligned}
\end{equation*}

\noindent Then:
\begin{equation*}
    \begin{aligned}
        & E(D|Z=z_m) - E(D|Z=z_{m-1}) \\ 
        &= E \left(\sum_{j=0}^J j \cdot I(D^{z_m} \geq j)|Z=z_m \right)-E \left(\sum_{j=0}^J j \cdot I(D^{z_{m-1}} \geq j)|Z=z_{m-1} \right) \\
        &= E \left(\sum_{j=0}^J j (I(D^{z_m}=j)-I(D^{z_{m-1}}=j)) \right) \\
        &= E \left(\sum_{j=0}^J j \cdot (I(D^{z_m} \geq j)-I(D^{z_m} \geq j+1) -I(D^{z_{m-1}} \geq j) - I(D^{z_{m-1}} \geq j+1)) \right) \\
        &= E \left(\sum_{j=1}^J I(D^{z_m} \geq j)-I(D^{z_{m-1}} \geq j) \right) \\
        &= \sum_{j=1}^J P(D^{z_m} \geq j > D^{z_{m-1}}).
    \end{aligned}
\end{equation*}

\noindent It is required that $P(D^{z_m} \geq j > D^{z_{m-1}})>0$ for some $j$ which imposes a relevance assumption on the instrument. \\ 

\noindent Plugging the above results into $\beta_{m,m-1}$, we get:
\begin{align*}
    \beta_{m,m-1} &= \frac{E(Y|Z=z_m)-E(Y|Z=z_{m-1})}{E(D|Z=z_m)-E(D|Z=z_{m-1})} \\
    &= \sum_{j=1}^J \frac{P(D^{z_m}\geq j > D^{z_{m-1}})}{\sum_{i=1}^J P(D^{z_m} \geq i > D^{z_{m-1}})} \cdot E(Y^j - Y^{j-1} | D^{z_m}\geq j > D^{z_{m-1}})\\
    &\hphantom{=} - \sum_{j=1}^J \frac{P(D^{z_{m-1}} \geq j > D^{z_m})}{\sum_{i=1}^J P(D^{z_m} \geq i > D^{z_{m-1}})} \cdot E(Y^{j-1} - Y^j| D^{z_{m-1}} \geq j > D^{z_m} ).
\end{align*}

\noindent Then, Equation (\ref{eq:theorem2_ia1994}) without imposing any monotonicity can be re-written to:
\begin{equation*}
    \begin{aligned}
        \beta_{\text{TSLS,PM}} &\equiv \sum_{m=1}^M \Bigl\{ I(D^{z_m}>D^{z_{m-1}}) \cdot \delta_{m,m-1} \cdot \omega_m \cdot \beta_{m,m-1}^c  \\
        & \qquad \quad - I(D^{z_m} < D^{z_{m-1}}) \cdot \delta_{m,m-1} \cdot \omega_m \cdot \beta_{m,m-1}^d \Bigr\},
    \end{aligned}
\end{equation*}
where 
\begin{equation*}
    \delta_{m,m-1} = E(D|Z=z_m) - E(D|Z=z_{m-1}),
\end{equation*}
and
\begin{equation*}
        \omega_m = \frac{\sum_{l=m}^M P(Z=z_l) (E(D|Z=z_l)-E(D))}{\sum_{l=0}^M P(Z=z_l) E(D|Z=z_l)(E(D|Z=z_l)-E(D))},
\end{equation*}
and 
\begin{equation*}
        \beta_{m,m-1}^c 
        = \sum_{j=1}^J \frac{P(D^{z_m} \geq j > D^{z_{m-1}})}{\sum_{i=1}^J P(D^{z_m} \geq i > D^{z_{m-1}})} E(Y^j - Y^{j-1} | D^{z_m} \geq j > D^{z_{m-1}}),
\end{equation*}
and 
\begin{equation*}
        \beta_{m,m-1}^d 
        = \sum_{j=1}^J \frac{P(D^{z_{m-1}} \geq j > D^{z_m})}{\sum_{i=1}^J P(D^{z_m} \geq i > D^{z_{m-1}})} \cdot E(Y^j - Y^{j-1}| D^{z_{m-1}} \geq j > D^{z_m} ),
\end{equation*}
where the superscripts $c$ and $d$ denote whether $\beta$ gives the LATE for those who respond as compliers or defiers for a change from $m-1$ to $m$ respectively.

The weights $\omega_m$ are equivalent to the presentation of the previous section. The weights $\delta_{m,m-1} \cdot \omega_m$ sum to one ($\sum_{m=1}^M\delta_{m,m-1} \cdot \omega_m=1$), and are non-negative ($\delta_{m,m-1} \cdot \omega_m > 0$ for all $m$). The weights are proportional to the impact that the instrument with $k$ used in constructing $\beta_{k,k-1}$ has on the treatment level. Similar to the previous section, more weight is given to $E(Y^j - Y^{j-1} | D^{z_m} \geq j > D^{z_{m-1}})$ and $E(Y^j - Y^{j-1}| D^{z_{m-1}} \geq j > D^{z_m} )$ if it lies in the center of the instrument distribution. 

\bigskip

\subsection{Proof for a continuous treatment}
\label{app:proof_continuous_treatment}

Combining the arguments in the present study with those of \citet{angrist2000interpretation}, the following can be shown:
\begin{align*}
    & E(Y|Z_1=Z_2=...=Z_K=1) - E(Y|Z_1=Z_2=...=Z_K=0) \\
    & = E(Y^{D^{1...1...1}}-Y^{D^{0...0...0}}) \\
    & = E \left( \int_0^{D^{1...1...1}}\frac{\partial Y^{t}}{\partial t} dt  - \int_{D^{0...0...0}}^{\infty}\frac{\partial Y^{t}}{\partial t} dt \right) \\
    & = E \left( \int_{D^{0...0...0}}^{D^{1...1...1}}\frac{\partial Y^{t}}{\partial t} dt \right) \\
    & = E \left( \int_{0}^{\infty} I\{D^{1...1...1} \geq t > D^{0...0...0} \} \frac{\partial Y^{t}}{\partial t} dt \right) \\
    & = \int_{0}^{\infty} E \left(  I\{D^{1...1...1} \geq t > D^{0...0...0} \} \frac{\partial Y^{t}}{\partial t}  \right) dt \\
    & = \int_{0}^{\infty} P(D^{1...1...1} \geq t > D^{0...0...0}) \cdot \frac{\partial  E(Y^{t}| D^{1...1...1} \geq t > D^{0...0...0})}{\partial t} dt .
\end{align*}

\noindent The independence assumption and the fundamental theorem of calculus ($f(x)=\int_0^x f'(t) dt = \int_0^x \frac{\partial f(t)}{\partial t} dt$) were used in lines two and three, respectively. Similarly, it can be shown that
\begin{align*}
    & E(D|Z_1=Z_2=...=Z_K=1) - E(D|Z_1=Z_2=...=Z_K=0) \\
    & = \int_{0}^{\infty} P(D^{1...1...1} \geq j > D^{0...0...0})  dj.
\end{align*}

\noindent Then:
\begin{align*}
    & \beta_{\text{CC-ACR}} = \frac{E(Y|Z_1=Z_2=...=Z_K=1) - E(Y|Z_1=Z_2=...=Z_K=0)}{E(D|Z_1=Z_2=...=Z_K=1) - E(D|Z_1=Z_2=...=Z_K=0)}\\
    & = \frac{\int_{0}^{\infty} P(D^{1...1...1} \geq t > D^{0...0...0}) \cdot \frac{\partial  E(Y^{t}| D^{1...1...1} \geq t > D^{0...0...0})}{\partial t} dt}{\int_{0}^{\infty} P(D^{1...1...1} \geq j > D^{0...0...0})  dj}.
\end{align*}

\bigskip
\section{LiM test}

\subsection{Inequalities for detecting local violations of LiM}
\label{app:inequalities_lim}

This section shows how the inequalities for the LiM test can be derived.
As the LiM assumption provides the condition that the CDFs do not cross, which is equivalent to the condition of having positive weights at every point of the distribution of $D$, the following $J+1$ inequalities have to hold under LiM (which can be derived from Equation (\ref{eq:weights})):
\begin{equation}\label{eq:raw_inequality_for_testing}
    P(D < j | Z_1=Z_2=...=Z_K=0) - P(D < j| Z_1=Z_2=...=Z_K=1) \geq 0, 
\end{equation}
for all $j \in \{0,1,...,J \}$. 

The inequalities in Condition (\ref{eq:raw_inequality_for_testing}) translate to learning the sign of the causal effect on the treatment variable of the sole instrument, $\widetilde{Z}=Z_1=Z_2=...=Z_K$, in the subsample of observations at the outer support of the instrument values:
\begin{equation*}
    P(D < j | \widetilde{Z}=0) - P(D < j|\widetilde{Z}=1) \geq 0 \text{ for all } j \in \{0,1,...,J \}.
\end{equation*}
Rewrite the previous equation to the following expression:
\begin{equation*}
    E(I(D < j) | \widetilde{Z}=0) - E(I(D < j)|\widetilde{Z}=1) \geq 0 \text{ for all } j \in \{0,1,...,J \}.
\end{equation*}
Then, the following inequality must be satisfied at any point $x$ in the covariate space:
\begin{equation*}
    E(I(D < j) | \widetilde{Z}=0, X=x) - E(I(D < j)|\widetilde{Z}=1, X=x) \geq 0 \text{ for all } j \in \{0,1,...,J \}.
\end{equation*}

\bigskip

\subsection{Test procedure}
\label{app:cftest_procedure}

The procedure by \citet{farbmacher2022instrument} can be followed for estimating $\tau_{j}(x)$ of Equation (\ref{eq:cf_tau}) and is described here. The average treatment effect given by this equation gives an insight into the magnitude of possible violations.
Two additional assumptions are required to establish causality: (1) $(Q_{J}^1,Q_{J}^0) \perp \widetilde{Z}|X$, and (2) $\epsilon < P(\widetilde{Z}=1|X=x)<1-\epsilon$ for some $\epsilon>0$.
Then, the average treatment effect can be estimated using augmented inverse-propensity weighting based on \citet{robins1994estimation}:
\begin{equation}\label{eq:gamma_test}
    \begin{aligned}
        \hat{\Gamma}_{j,i} \equiv &\hat{\tau}_{j}^{(-i)}(X_i) \\
        &+ \frac{\widetilde{Z}_i - \hat{e}^{(-i)}(X_i)}{\hat{e}^{(-i)}(X_i) \left( 1 - \hat{e}^{(-i)}(X_i) \right)} 
        \times \left( Q_{j,i} - \hat{\mu}_{j}^{(-i)}(X_i) - (\widetilde{Z}_i-\hat{e}^{(-i)}(X_i)) \hat{\tau}_{j}^{(-i)}(X_i)  \right).
    \end{aligned}
\end{equation}
$\hat{\tau}_{j}(X_i)$,  $\hat{e}(X_i)$, and  $\hat{\mu}_{j}(X_i)$ are estimates of $\tau_{j}(x)$, $e(x)=P(\widetilde{Z}_i=1|X_i=x)$, and $\mu_{j}(x)=E(Q_{j,i}|X_i=x)$, respectively. The superscript $(-i)$ denotes out-of-bag estimates. This means that estimates were obtained without the $i$th observation (e.g., $D_i$ did not contribute to estimating $\hat{\tau}_{j}^{(-i)}(X_i)$). 

The full sample is randomly split into two samples, $S^A$ and $S^B$, each of which will be used both for training and predicting. Denote the trees resulting from these samples for each value of $j$ by $\Pi_{j}^{S{^A}}$ and $\Pi_{j}^{S{^B}}$.   
Then consider the expectation of $\Gamma_{j,i}$ for a given partition
\begin{equation*}
    \zeta_{j,l}^A=E\left( \Gamma_{j,i} | X_i \in L_l \left(x;\Pi_{j}^{S{^B}} \right) \right),
\end{equation*}
\begin{equation*}
    \zeta_{j,l}^B=E\left( \Gamma_{j,i} | X_i \in L_l \left(x;\Pi_{j}^{S{^A}} \right) \right).
\end{equation*}
Let $L_l \left(x;\Pi_{j} \right)$ denote the $l$th element of the collection of leaves of the tree $\Pi_{j}$. The moments of all leaves are contained in $\zeta = \left( \zeta^A,\zeta^B \right)$. Recall that positive values of $\zeta$ point toward a local violation of LiM. Then a local violation of LiM can be tested with the following hypothesis test:
\begin{equation*}
    \begin{aligned}
        & H_0: \zeta_s \leq 0 \text{ \indent for all } s=1,...,p \\
        & H_1: \zeta_s > 0 \text{ \indent for some } s=1,...,p,
    \end{aligned}
\end{equation*}
where $p=|\zeta|$ is the number of sample splits. This means that $p=2$, when splitting the sample into two samples, $S^A$ and $S^B$.

Under the null hypothesis, an upper bound on the $(1-\alpha)$ quantile of $\sqrt{n} \left(\hat{\zeta}_j - \zeta_j \right)/\hat{\sigma}_j$ is enough for testing.:
\begin{equation*}
    T = \max_{1 \leq s \leq p} \frac{\sqrt{n} \hat{\zeta}_j}{\hat{\sigma}_j} \leq \max_{1 \leq s \leq p} \frac{\sqrt{n} \left(\hat{\zeta}_j - \zeta_j \right)}{\hat{\sigma}_j}.
\end{equation*}
Finally, the p-values should be Bonferroni corrected for multiple hypothesis testing.

Asymptotic results can be derived as in \citet{farbmacher2022instrument} using the results of \citet{chernozhukov2018double}.

\bigskip
\newpage
\subsection{Pseudo code of the LiM testing procedure}
\label{app:cftest_pseudo_code}

I build upon the procedure and code of \citet{farbmacher2022instrument} to establish a test for LiM. The pseudo code for the LiM test procedure is presented by Algorithm \ref{alg:cf_test}.  \bigskip

\begin{algorithm}
\caption{LiMtest}\label{alg:cf_test}
Input: $n_s$ observations $(D_i,\widetilde{Z}_i,X_i)$ with $D_i \in \{ 0,1,...,J \}$ the treatment, $\widetilde{Z}_i$ the instrument indicator for the outer support of the instrument distribution, and $X_i$ the covariates. The minimum leaf size is denoted $k$, and the significance level with $\alpha$.
\begin{algorithmic}[1]
\For{$j=0,1,...,J$} 
    \State Construct the pseudo variable $Q_{j,i}$.
    \For{both samples separately} 
        \State \parbox[t]{\dimexpr\linewidth-\algorithmicindent}{%
        Obtain leave-one-out estimates $\hat{\mu}_j^{(-i)}(X_i)$ with a regression forest using outcome $Q_{j,i}$ \\ \vspace{0.2cm}and including covariates $X_i$.}
        \State \parbox[t]{\dimexpr\linewidth-\algorithmicindent}{%
        Obtain leave-one-out estimates $\hat{\tau}_j^{(-i)}(X_i)$ with a causal forest using outcome $Q_{j,i}$ and \\ \vspace{0.2cm}including covariates $X_i$.}
        \State Construct the estimates $\hat{\Gamma}_{j,i}$ as in Equation (\ref{eq:gamma_test}) in Appendix \ref{app:cftest_procedure}.
    \EndFor 
    \State \parbox[t]{\dimexpr\linewidth-\algorithmicindent}{%
    Fit a CART tree on sample $A$ using outcome $\hat{\Gamma}_{j,i}$, covariates $X_i$, minimal leaf size $k$, and apply cost complexity pruning.}
    \vspace{-0.1cm}
    \For{each leaf $l=1,...,l_{\text{max}}$}
        \State Calculate the t-statistic $t_{j,l}^{(A)}$ over units $\hat{\Gamma}_{j,i}$ in sample $A$ present in leaf $l$.
        \If{$t_{j,l}^{(A)}>\Phi^{-1}(1-0.05/l_{\text{max}})$}
            \State \parbox[t]{\dimexpr\linewidth-\algorithmicindent}{%
            Calculate the t-statistic $t_{j,l}^{(B)}$ over units $\hat{\Gamma}_{j,i}$ in sample $B$ present in leaf $l$ and store \\ \vspace{0.2cm}the values in a vector $T_{\text{vec}}$.} 
            \vspace{-0.2cm}
        \EndIf
    \EndFor
    \State Repeat lines 8-14 with the roles for samples $A$ and $B$ switched.
    \If{$\text{max}(T_{\text{vec}})>\Phi^{-1}(1-\alpha/|T_{\text{vec}}|)$}
        \State Reject the null hypothesis.
    \EndIf
\EndFor
\end{algorithmic}
\end{algorithm}

\bigskip
\newpage

\subsection{Simulation study - LiM test}
\label{app:simulation_limtest}

In this section, we evaluate the performance of the LiM test through a simulation study.
I conduct an empirical Monte Carlo study closely modeling the data from \citet{card1995geographic}. 
Specifically, I generate the instrumental variable $\widetilde{Z}$ using a binomial distribution with probabilities matching the mean observed.
The control variables are generated are generated using a multivariate normal distribution, with means and covariance matrix derived from the empirical data. After generating the continuous control variables, I binarize them based on the mean. 
The sample size is 1500.
For simplicity, I consider three different treatment levels, $D\in\{12,13,14\}$.
Moreover, I introduce different complier types with specific probabilities of occurrence (type shares), as shown in Table \ref{tab:simulation_setup_limtest}. I consider two scenarios: one where LiM is valid, meaning no combined defiers are present, and another where LiM is locally violated due to the presence of defier types in the southern region, who change their treatment level from 13 to 12 when both instruments equal one.
1,000 simulation repetitions are performed. 

Table \ref{tab:results_sim_limtest} reports the rejection rates. The test successfully detects the violation for the combined defiers in the Southern region at the correct treatment level.
The results reflect that the test is sensitive to the covariates included. As proxy variables might be chosen as most important variables for splitting, the tree structure with the maximum violation can only give an indication for the subgroup where LiM might be violated.

\begin{table}[bp]
    \centering
    \caption{This table presents an overview of the different response types and their respective shares. The simulation study considers two scenarios: one where LiM is valid, and another where LiM is locally violated due to the presence of combined defiers.}
    \label{tab:simulation_setup_limtest}
    \begin{tabular}{lccccc}
        \toprule
        & & & LiM valid & LiM violated \\
        \cmidrule{4-5} 
        & $D^{00}$ & $D^{11}$ & Type share & Type share \\
        \midrule
        Combined compliers & & & & \\
        \addlinespace 
        $cc_{12,13}$ & 12 & 13 & 15\% & 5\% \\
        $cc_{13,14}$ & 13 & 14 & 10\% & 10\% \\
        $cc_{12,14}$ & 12 & 14 & 5\% & 5\% \\
        \addlinespace \hline
        Combined non-responders & & & & \\
        \addlinespace
        $cn_{12,12}$ & 12 & 12 & 25\% & 25\% \\
        $cn_{13,13}$ & 13 & 13 & 20\% & 20\% \\
        $cn_{14,14}$ & 14 & 14 & 25\% & 25\% \\
        \addlinespace \hline
        Combined defiers & & & & \\
        \addlinespace
        $cd_{13,12}$ where $south=1$ & 13 & 12 & - & 10\% \\[0.5em] 
        \hline
        \addlinespace
        & & & 100\% & 100\% \\
        \bottomrule
    \end{tabular}
\end{table}

\begin{table}[btp]
    \centering
    \caption{This table presents the rejection results for the setting where LiM is valid and when LiM is violated. 
    }
    \label{tab:results_sim_limtest}
    \begin{tabular}{lccccc}
        \toprule
        & \multicolumn{2}{c}{LiM valid} & \multicolumn{2}{c}{LiM violated} \\
        \cmidrule(lr){2-3} \cmidrule(lr){4-5}
        & 12 to 13 & 13 to 14 & 12 to 13 & 13 to 14 \\
        \midrule
        Rejection rate & 0\% & 0\% & 100\% & 0\% \\ 
        First split at \textit{south} = 1 & - & - & 94.8\% & - \\ 
        \bottomrule
    \end{tabular}
\end{table}

\bigskip

\section{Details on the model specifications}
\label{app:model_specifications}

\subsection{Specifications of the machine learners employed in Section \ref{sec:estimation_results}}
\label{app:specifications_schooling}

Lasso (Least Absolute Shrinkage and Selection Operator) introduced by \citet{tibshirani1996regression} is a regularization technique that performs both variable selection and regularization to enhance prediction accuracy. It includes a penalty term proportional to the absolute value of the coefficients, which helps in shrinking some coefficients to zero and thus performing feature selection. For Lasso, a fully saturated specification including all covariate interactions is considered, while raw controls are included in the other methods.
The key hyperparameter is the regularization parameter, lambda. I use the \textit{glmnet} library (version 4.1-8) in R to perform 5-fold cross-validation to select the optimal lambda.

Random Forest is an ensemble learning method used for classification and regression, introduced by \citet{breiman2001random}. It operates by constructing multiple decision trees during training and outputting the mode of the classes (classification) or mean prediction (regression) of the individual trees. Key hyperparameters include the number of trees, minimum node size, and the number of variables sampled at each split.
The tuning process involved a grid search combined with 5-fold cross-validation.
Specifically, the grid for the number of trees contained the values 500 and 1000. For the minimum node size, I consider values of 25, 50, and 100. The number of variables sampled at each split is varied with values of 2, 3, and 4. I use the \textit{caret} package (version 6.0-94) and the \textit{randomForest} package (version 4.7-1.1) to streamline the model training and hyperparameter tuning process.

Boosted Trees, specifically Gradient Boosting, is another ensemble learning technique that builds models sequentially, originally introduced by \citet{friedman2001greedy}. Each new model attempts to correct errors made by the previous models. The \textit{gbm} library (version 2.1.9) in R is used for implementing Boosted Trees.
I conduct a grid search on several key hyperparameters: the interaction depth, number of trees, and shrinkage rate. The grid for the interaction depth includes values of 1, 2, and 3. The number of trees is varied with values of 100, 200, and 300. For the shrinkage (learning rate), I consider values of 0.1, 0.05, and 0.01. 5-fold cross-validation is used to select the optimal combination of hyperparameter values.

In tuning the hyper parameters, for treatment and outcome models, RMSE (Root Mean Squared Error) was used to evaluate the model performance.
Additionally, for each of the three methods (LASSO, Random Forest, and Boosted Trees), trimming of the predicted propensity scores was performed with a value of 0.01, such that they lie between 0.01 and 0.99.

\begin{table}[ht]
\centering
\caption{This table presents the mean RMSE over 25 splits for the machine learners used to obtain DML estimates in \citeauthor{card1995geographic}'s (\citeyear{card1995geographic}) study, as presented in Table \ref{tab:all_estimates_schooling}, for selecting the best estimator among different machine learners.}
\label{tab:rmse_ml_card}
\begin{tabular}{lccc}
  \hline
 & Mean RMSE for $y$ & Mean RMSE for $d$ & Mean RMSE for $\widetilde{Z}$  \\ 
  \hline
  Lasso & 0.42 & 1.96 & 0.37  \\ 
  Random forest & 0.41 & 1.93 & 0.36 \\ 
  Boosted trees & 0.41 & 1.92 & 0.36  \\  
   \hline
\end{tabular}
\end{table}

\bigskip

\subsection{Specifications of the machine learners employed in Section \ref{sec:estimation_results_twins}}
\label{app:specifications_twins}

The choices are similar to those outlined in Appendix \ref{app:specifications_schooling}, with two exceptions: third-order covariate interactions are included for Lasso, and for random forests, only 100 and 500 trees are considered due to computation time.

\begin{table}[ht]
\centering
\caption{This table presents the mean RMSE over 5 splits for the machine learners used to obtain DML estimates in \citeauthor{angrist1996children}'s (\citeyear{angrist1996children}) study, as presented in Table \ref{tab:all_estimates_children}, for selecting the best estimator among different machine learners.}
\label{tab:rmse_ml_children}
\begin{adjustbox}{max width=\textwidth}
\begin{tabular}{@{}lccc@{}}
\hline
 & Mean RMSE for $y$ & Mean RMSE for $d$ & Mean RMSE for $\widetilde{Z}$  \\ \hline
 \multicolumn{4}{l}{Panel A: Causal effect on labor income} \\ \hline
  Lasso & 4574.816 &  0.676 &  0.010  \\ 
  Random forest &  4542.167 & 0.679  & 0.108 \\ 
  Boosted trees & 4525.689  & 0.672 & 0.021  \\  
   \hline
\multicolumn{4}{l}{Panel B: Causal effect on hours worked per week} \\ \hline
  Lasso &  17.853 & 0.676 &  0.010  \\ 
  Random forest &  17.855 & 0.679  & 0.108  \\ 
  Boosted trees &  17.799 & 0.672 &  0.021 \\  
   \hline
\multicolumn{4}{l}{Panel C: Causal effect on weeks worked} \\ \hline
  Lasso &  21.30 & 0.68 & 0.01  \\ 
  Random forest & 21.33 & 0.68 & 0.11   \\ 
  Boosted trees & 21.25 & 0.67 & 0.02   \\  
   \hline
\end{tabular}
\end{adjustbox}
\end{table}

\bigskip

\end{appendices}

\end{document}